\documentclass[a4paper,11pt]{article}
\usepackage{jcappub} 
\usepackage{lineno}
\usepackage{color}

\title{\boldmath Numerical simulation of common envelope stage in close binary systems: comparison between Newtonian and modified gravity}

\author{A.V. Astashenok$^{+}$, A.S. Baigashov, A.S. Tepliakov, K.P. Gusev, E.R. Shamardina}
\affiliation{Immanuel Kant Baltic Federal University,\\
 Department of Physics, Technology and IT,\\
236041 Kaliningrad, Nevskogo str. 14, Russia}

\emailAdd{$^{+}$artyom.art@gmail.com}

\abstract{We consider the important stage in evolution of close binary system namely common envelope phase in framework of various models of modified gravity. The comparison of results between calculations in Newtonian gravity and modified gravity allows to estimate possible observational imprints of modified gravity. Although declination from Newtonian gravity should be negligible we can propose that due to the long times some new effects can appear. We use the moving-mesh code AREPO for numerical simulation of binary system consisting of $\sim M_{\odot}$ white dwarf and a red giant with mass $\sim 2 M_{\odot}$.
For implementing modified gravity into AREPO code we apply the method of (pseudo)potential, assuming that modified gravity can be described by small corrections to usual Newtonian gravitational potential. As in Newtonian case initial orbit has to shrink due to the energy transfer to the envelope of a giant. {We investigated evolution of common envelope in a case of simple model of modified gravity with various values of parameters and compared results with simulation in frames of Newtonian gravity.} }

\begin{document}
\maketitle

\flushbottom

\section{Introduction}

The details of common envelope (CE) phase \textcolor{blue}{in the evolution} of close binaries are not completely understand from theoretical viewpoint till now. According to generally accepted picture the matter from giant star (primary component) starts to flow on secondary less massive component (white dwarf (WD) or solar-type star). In result of tidal forces or hydrodynamical instabilities of mass flow the secondary star plunges into the envelope of the \textcolor{blue}{giant star}. Due to the frictional drag orbit of the secondary components shrinks and it moves toward the core of the giant star. The final result of this sufficiently fast process \textcolor{blue}{is merging} of the components or envelope ejecting. 

Detailed description of evolution of common envelope is very important for understanding the appearance of the progenitors of Type Ia supernovae, X-ray binaries, double neutron stars and potential sources of gravitational waves. 

The results of various hydro-dynamical calculations of common envelope evolution have been presented in literature. Authors of \cite{Ivanova_2013} compared the results of hydrodynamic simulations from different groups and discussed the potential effect of initial conditions on the differences in the outcomes. The main result is that the problem is very complex for both numerical calculations and analytic treatment. A relatively common problem would be one in which a neutron star or RG spirals into the envelope of a \textcolor{blue}{red giant (RG)}. Simulations of such a CE event might need to cover a range in time scale of 10$^{10}$ (i.e. from 1 s, which is already three orders of magnitude longer than the dynamical time scale of the \textcolor{blue}{neutron star} (NS), to 1000 yr, the thermal time of the envelope and plausible duration of the common envelope phase). The corresponding range of spatial scales could be $\sim 10^8$ (i.e. from 10 km, the size of the NS, to $\sim$ 1000 $R_{\odot}$ ).

In \cite{Iaconi_2019} almost all existing 3D hydrodynamic simulations and observations of single degenerate, post-CE binaries are considered. Also authors investigated the proposition that final separation between components is defined by fraction of released orbital energy that is used to do work and unbind the envelope gas. There is no clarity about the effects of the spatial resolution and the softening length on the results.

The simulation of low mass binary ($1.05 M_{\odot}$ RG and $0.6M_{\odot}$ companion) is described in \cite{Ricker_2012}. After a fast inspiral phase one quarter of envelope is ejected from the system and mass continues to be lost. The initial orbital separation decreases sevenfold due to loss of angular momentum and energy. 

Comprehensive simulations in \textcolor{blue}{frames of smoothed-particle hydrodynamics} (SPH) and uniform grid codes were performed in \cite{Passy_2012}. The mass of red giant branch star was $0.88 M_{\odot}$ and the mass of companion star varied from 0.9$M_{\odot}$ to $0.1M_{\odot}$. It is interesting to note that in these simulations envelope was not able to be ejected and the final orbital separations were much larger than observed in post-CE systems.

The authors of \cite{Ohlmann_2016} used AREPO code with moving mesh for studying evolution of common envelope for $\sim 2 M_\odot$ RG (possessing 0.4$M_{\odot}$) \textcolor{blue}{and a $1 M_{\odot}$} RG. AREPO combines advantages of smoothed particle hydrodynamic and traditional grid-based hydrodynamic codes. After some time a new phenomenon is observed. Large-scale flow instabilities are triggered by shear flows between adjacent shock layers. These indicate the onset of turbulent convection in the envelope, thus altering the transfer of energy on longer timescales. However at the end of simulation, only 8\% of the envelope mass is ejected from system.

The same result takes place also for \textcolor{blue}{star on asymptotic giant branch (AGB)}, although the envelope of such star is less tightly bound than that of a RG (see \cite{Sand_2020}). Only one fifth of envelope mass is ejected. As the authors proposed, taking into account ionization energy can lead to complete envelope ejection.

{It is interesting to consider evolution of common envelope assuming non-Newtonian gravity and pose the question about various imprints of modified gravity in evolution. In principle we can propose that near a compact object (RG or neutron star) gravitational field differs from Newtonian potential.} 

Initially, models of extended gravity are motivated by cosmology. 
Accelerated expansion of universe \cite{Riess-1,Perlmutter,Riess-2} is usually explained in frames of $\Lambda $CDM
model. According to this model, dark energy is constant
vacuum energy with density consisting of around 70\%
of all energy density in the universe. Usual
baryon matter provides only 4 \%. The rest is so-called dark
matter. Although this model describes observational data, some unresolved issues remain. Firstly, the value of cosmological constant is very small in comparison to one predicted by quantum field theory. Another problem is the nature of dark matter, which constitutes of nearly 25\% of universe.  

The possible explanation of accelerated cosmological expansion can be given in various models of extended gravity
(see \cite{Odintsov1}, \cite{Turner}).
Modified gravity theories also can provide a unified description of
cosmological evolution including early inflation, matter and
radiation dominance era in $f(R)$ theory \cite{Nojiri-5},
\cite{Nojiri-4}.

Deviation of gravity from General Relativity can lead to some consequences at the astrophysical level. The gravitational field near the compact stars is extremely strong and
therefore the question about possible deviations from General
Relativity arises. Compact stars (especially neutron stars) in simple models of modified gravity have been extensively investigated in multiple works (for a comprehensive review of
compact star models in modified theories of gravity see
\cite{Olmo} and references therein). In particular some interesting results were obtained for \textcolor{blue}{models of neutron stars in $R^2$ gravity}. Scalar curvature quickly
drops outside the surface and from some distance one can assume that $R=0$ and
therefore we have a solution corresponding to Schwarzschild
solution with some mass $M_s$, where $M_s$ is not equal to mass
confined by star surface. From the viewpoint of a distant observer,
mass $M_s$ is gravitational mass of a neutron star and this mass is larger than the gravitational mass enclosed within star surface.

In this paper, we consider evolution of RG envelope in close binary system consist of a RG and a compact RG. We assume that the field of the RG contains some corrections due to modified gravity. In the next section, we provide a brief description of the RG model, which was used in our calculations. Then, we consider a simple model of modified gravity and the corresponding spherical symmetry solution for gravitational field outside the compact star. In order to implement these solutions into AREPO code we use formalism of pseudopotential, i.e. we add some correction to Newtonian gravitational force. The fourth section is devoted to the results of the numerical  calculations of common envelope evolution within the framework of Newtonian gravity, as well as the consideration of the model of modified gravity for some values of parameters.  
    
\section{The stable model of red giant in simulations and binary system setup}

\subsection{Red giant in MESA}

As part of our study, in order to model the structure of the RG we \textcolor{blue}{used} the MESA (Modules for Experiments in Stellar Astrophysics) toolbox, which is a comprehensive open-source library covering a wide range of astrophysical calculations \cite{Paxton_2011,Paxton_2013,Paxton_2015,Paxton_2018}. MESA integrates numerous specialized modules, each designed to solve a specific class of problems, including equations of state, material opacity, nuclear reaction rates, element diffusion parameters, and conditions at the boundary of a star's atmosphere.

\begin{figure}
\centering
    \includegraphics[width=0.75\textwidth]{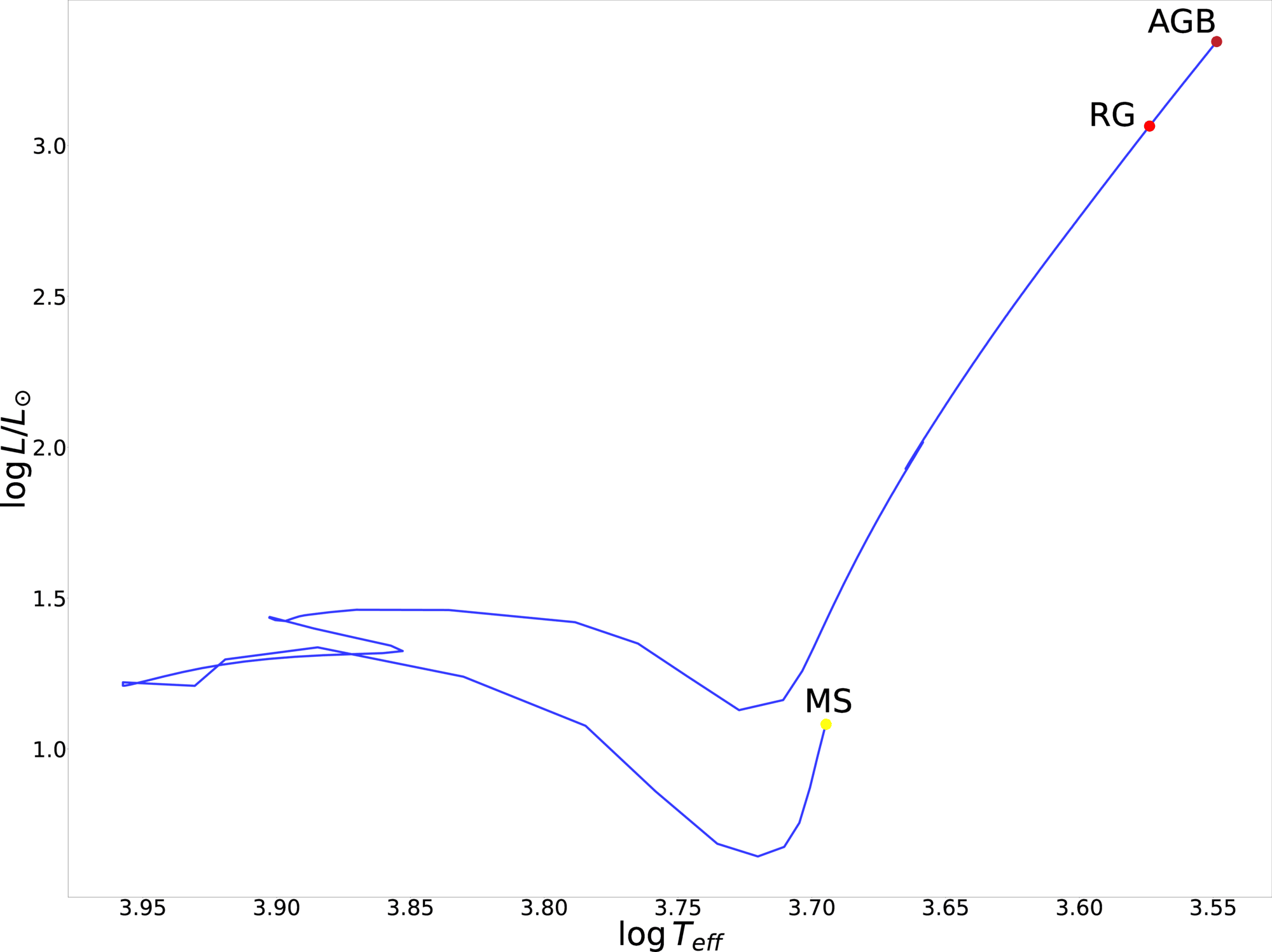}
    \caption{Hertzsprung–Russell diagram for the evolution of a 2$M_{\odot}$ star using MESA. Three stages of star development are identified: main sequence (MS), RG and star on AGB.} 
    \label{Hertzsprung–Russell diagram}
\end{figure}

We start from a main-sequence star with mass $2M_{\odot}$ and with metallicity $z = 0.02$, evolving up to the AGB stage. Figure \ref{Hertzsprung–Russell diagram} shows the evolutionary trajectory according to the Hertzsprung–Russell diagram, which also shows the initial stage of the star’s development, the RG stage selected for subsequent analysis, and the final stage of evolution.

In astrophysical research, when modeling complex processes such as the dynamic behavior of stars, their interactions, mergers and convective motions, one-dimensional codes are insufficient to achieve realistic representation. These processes require a more complex approach, which can be implemented via 3D modeling. This approach allows us to take into account multiple  factors and interactions that cannot be adequately reproduced within one-dimensional models.

One-dimensional density profile obtained using MESA can serve as a basis for creating the initial conditions in three-dimensional hydrodynamic models of red giants \cite{Ohlmann_2017}. This is achieved by projecting a one-dimensional spherically symmetrical profile derived from the stellar evolution code onto a three-dimensional grid used in hydrodynamic calculations. However, the central  density of a star can significantly exceed the density of the outerlayers, which leads to complications when modeling density gradients using particles of the same mass. To accurately represent these gradients requires a colossal number of particles, rendering the task nearly impossible. The inner layers of the star can be represented by a single particle, called the "core", whose mass is equal to the difference between the total mass of the star and the mass represented in the particle distribution. This approach can significantly reduce the number of particles needed for the simulation while maintaining an accurate representation of the outer layers of the star. 

For conversion of one-dimensional profile to 3D model of star we used $^{1D}$MESA2\-HYDRO$^{3D}$ \cite{Joyce_2019}, a Python-based open source tool. Key physical parameters required for $^{1D}$MESA2HYDRO$^{3D}$ include mass, density, pressure and internal energy, expressed as a function of radius. The transformation of one-dimensional data into three-dimensional space is accomplished by translating discrete values of radius $r$ and density $\rho$ into a set of mass and radial coordinates, which are interpreted as a distribution of particles. These coordinates are denoted as $N$ and $R$, where $N$ is the “number” of particles and $R$ is the “radius”. The radial coordinate $R$ is defined as the average value between the inner and outer boundaries of the spherical shell containing the mass, and used to create a distribution of particles covering the surface of a sphere. The coordinate $N$ corresponds to the integer used by the Hierarchical Equal Area isoLatitude Pixelization (HEALPix) spherical tessellation algorithm \cite{Gorski_2005}, which is used to calculate the total number of particles $n_{p}$ uniformly distributed over the spherical shell. This surface distribution is coupled with stacking technique proposed in \cite{Pakmor_2012}. The strength of this distribution method is that it provides both smooth and random \textcolor{blue}{initial conditions} and thus minimizes the occurrence of nonphysical artifacts during system evolution.

For our of RG model we use the radial cut-off as \textcolor{blue}{0.05$R_0$ where $R_0$} is initial radius ($\sim 46 R_\odot$). This part of the star is replaced by point mass ($\sim0.395M_{\odot}$). The \textcolor{blue}{remaining mass} ($\sim 1.605 M_{\odot}$) consists of $\sim 1.1\times 10^{6}$ particles.

\subsection{Relaxation of red giant model}

Due to the differences in discretization of pressure and gravity in the numerical schemes matching spherically symmetric stellar models with multidimensional hydrodynamic grids often leads to disruption of hydrostatic equilibrium. Pressure is taken into account in the flow calculations within a finite volume scheme, while gravity is calculated pointwise using a tree method. Additional errors arise when interpolating a high-resolution star profile onto a coarser hydrodynamic grid, which leads to a violation of hydrostatic equilibrium and the appearance of false velocities.

To eliminate these velocities, a relaxation procedure is applied, which allows obtaining stable models in hydrodynamic simulations on several dynamic time scales. We used a scheme based on \cite{Pakmor_2012}, \cite{Rosswog_2004}, \cite{Ohlmann_2017}, which has being proven to be effective in creating stable models. To damp out velocity fluctuations, we add to dynamical equations the friction-like term
$$
\dot{\vec{v}} = -\frac{1}{\tau}\vec{v}
$$
At the start of the modeling process, the parameter $\tau$ is chosen to be small ($0.1t_{dyn}$), where for the dynamic time scale $t_{dyn}$ we used the sound transit time. Then $\tau$ increases to $t_{dyn}$, which corresponds to a decrease in attenuation, according to the following formulas:
\begin{equation}
  \tau(t) = \begin{cases}
    \tau_1, \quad & t < 2t_\mathrm{dyn} \\
    \tau_1 \left( \frac{\tau_2}{\tau_1} 
    \right)^{\frac{t-2t_\mathrm{dyn}}{3t_\mathrm{dyn}}}, \quad & 2t_\mathrm{dyn}
    < t < 5t_\mathrm{dyn} \\
    \infty, \quad & t > 5t_\mathrm{dyn} 
  \end{cases}.
  \label{eq:relaxationtime}
\end{equation}

The core of the giant is replaced by a
point mass that only gravitationally interacts with the envelope.
For gravitation-only particles gravitational acceleration in AREPO is given by
a \textcolor{blue}{piecewise} function \cite{Springel_2010}
\begin{equation}
  g_\mathrm{c}(r) = - G m_\mathrm{c} \frac{r}{h^3} 
  \begin{cases}
    -\frac{32}{3} + u^2 \left( \frac{192}{5} - 32 u \right), & 0 \leq u <
    \frac{1}{2}, \\
    \frac{1}{15 u^3} - \frac{64}{3} + 48 u \\ - \frac{192}{5}u^2 +
    \frac{32}{3} u^3, & \frac{1}{2} \leq u < 1, \\
    - \frac{1}{u^3},  & u \geq 1,
  \end{cases}
  \label{eq:spline}
\end{equation}
where $u = r/h$, $h$ is the softening length of the interaction, and
$m_\mathrm{c}$ is the mass of the particle representing the core.

\begin{figure}
\centering
    \includegraphics[width=0.75\textwidth]{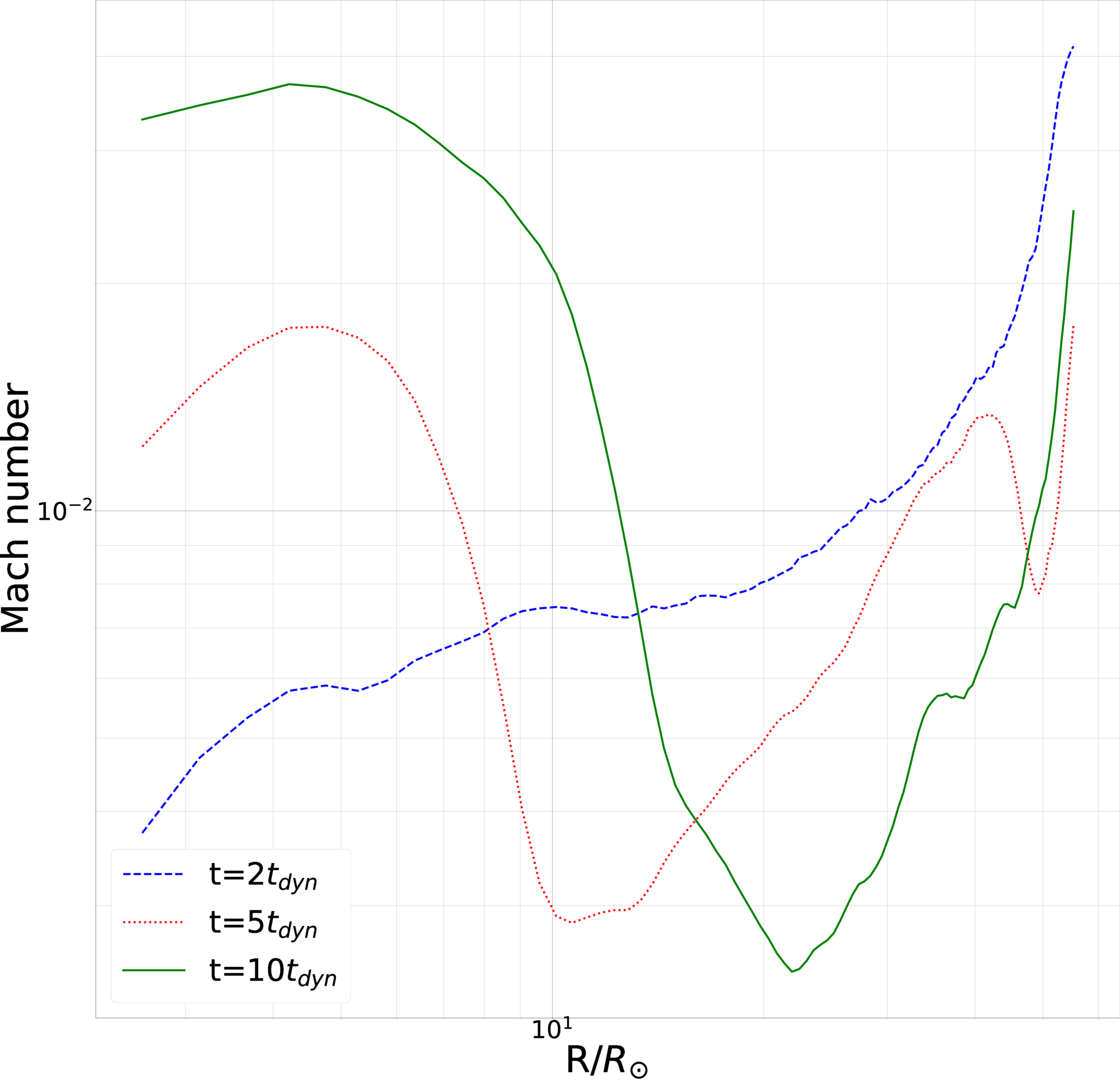}
    \caption{Dependence of radially binned and averaged Mach number from the radial coordinate for RG at different times of relaxation run ($t=2t_{dyn}, 5t_{dyn}, 10t_{dyn}$).  }    
    \label{Mach_Radius}
\end{figure}

\begin{figure}
\centering
    \includegraphics[width=0.5\textwidth]{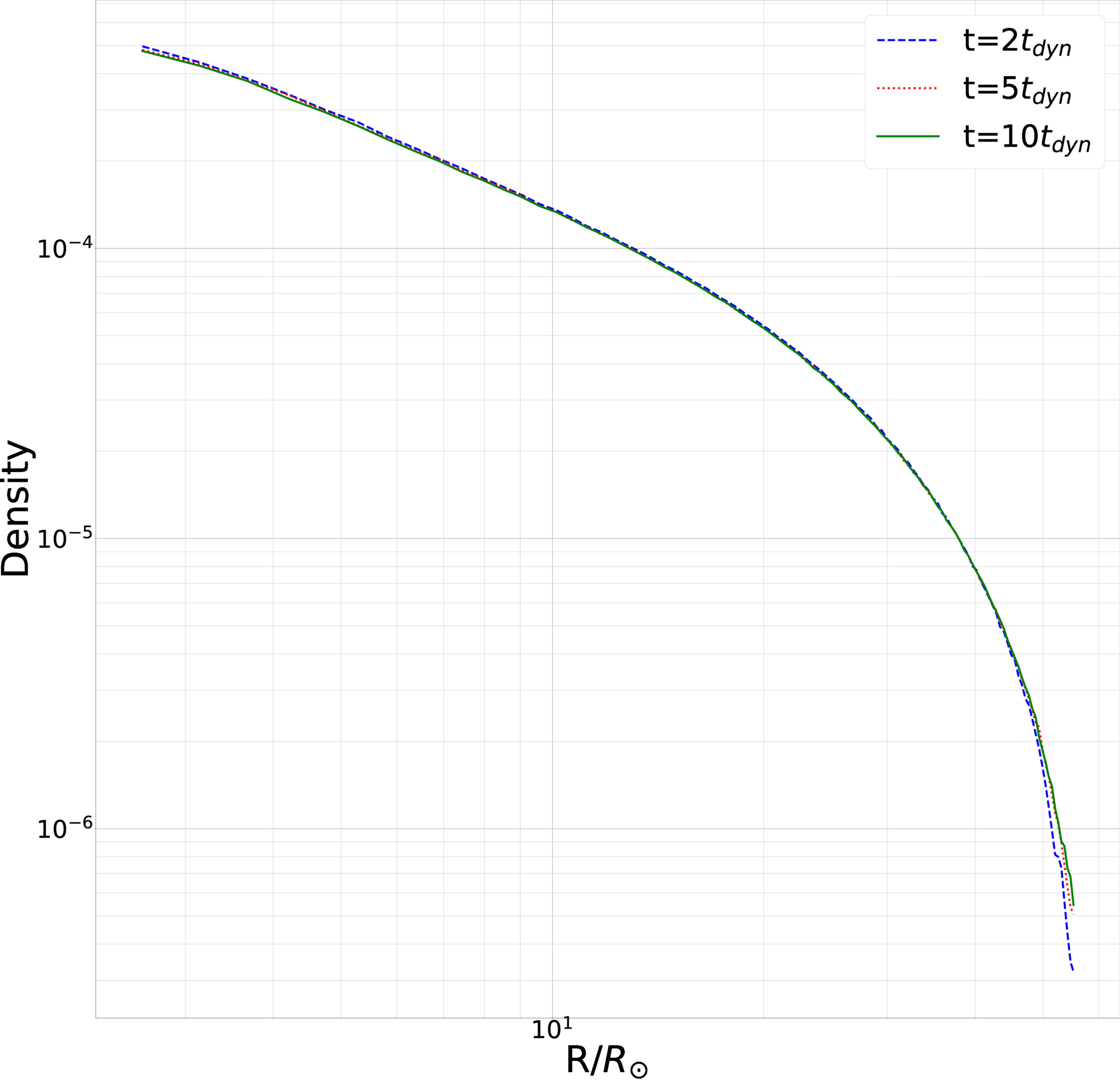}\includegraphics[width=0.5\textwidth]{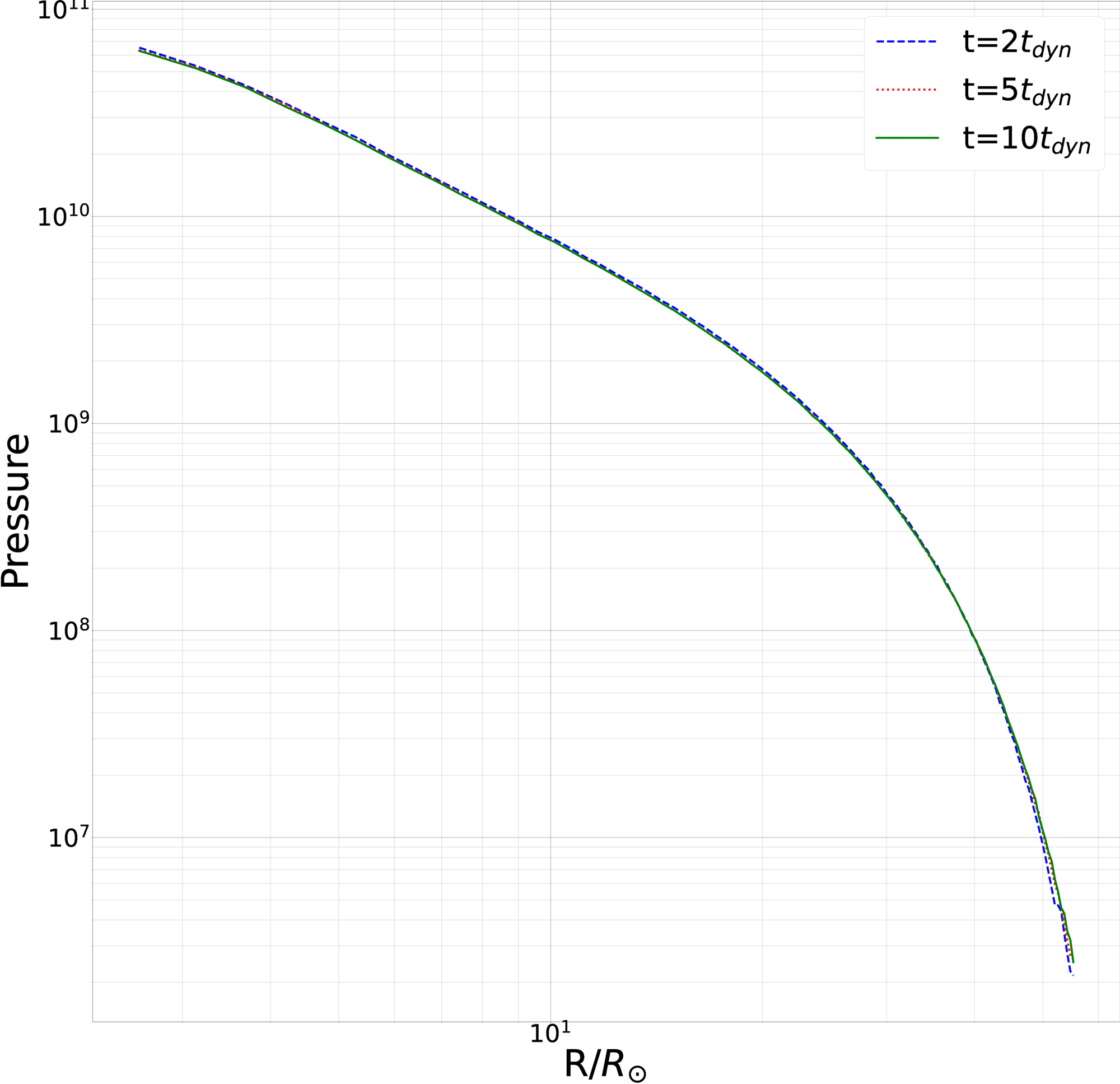}
    \caption{Dependence of energy density and pressure on the radial coordinate for RG at different times of relaxation run ($2t_{dyn}, 5t_{dyn}, 10t_{dyn}$).}  
     \label{Rho_Radius}
\end{figure}

After $t=5t_{dyn}$, we get in the result of numerical simulations  we obtained a stable 3D model of the star with an envelope mass of {1.575 $M_{\odot}$}, giving a total stellar mass of {1.97} $M_{\odot}$. The Mach numbers in the outer layers of the star are around $10^{-3} - 10^{-2}$, (see Fig. \ref{Mach_Radius}), which is consistent with expectations based on the original MESA model. We have also checked the stability of density and pressure profiles over the course of the simulation, and these profiles remain stable after completion of the relaxation procedure, when damping term is turning off (Fig. \ref{Rho_Radius}). Thus, the use of the relaxation procedure makes it possible to minimize sampling errors and obtain reliable hydrodynamic models of stars, which is a significant step in the field of modeling stellar dynamics and evolution.

\subsection{Binary system of a red giant and a white dwarf}

We model the phase of the common envelope of a binary system consisting of a RG and a WD. This stage of evolution takes place when the outer layers of the RG begin to fall on the second component, during which the core of the RG and the WD enter a fast inspiral (see \cite{Ohlmann_2016}). In this process, part of the gas of the RG shell ejects into space, and the rest of it forms a common spherical envelope with a core in the form of a close binary system of two white dwarfs. Thanks to the finite volume hydrodynamics moving-mesh code AREPO, this numerical experiment can be carried out with great accuracy, which will not only determine the GR deviations from the Newtonian gravity, but will also result in more precise testing of the parameters of modified gravity considered below.

\textcolor{blue}{We can expect that due to relatively small densities of giant stars the corresponding field in center of such star and around is very close to Newtonian and we don't take into account corrections from general relativity (GR) or modified gravity (MG) in field of RG.} In such conditions, we consider the evolution of the phase of the CE of the binary system, there a large contribution to the accretion of the substance of the RG to the WD is made by the GR and MG, near the WD. The core of the RG is a gravitation-only particle similar to a companion - a WD. As a result, the model consists of two  gravitation-only particles (a WD and a RG nucleus - “core cells”), a gas atmosphere of a RG and a background gas (“gas cells”) that fills the entire space.

We place the RG in the center of the coordinate system and we place the RG so that the RG is close to surface of the RG (this corresponds to the initial distance between RG core and WD $\sim 57.5 R_{\odot}$). Initial x-coordinate of RG is positive and y-coordinate is zero; the components of the binary system rotate in (x,y)-plane clockwise.  Velocities are taken as corresponding to the circular orbits of the components. For the Voronoi mesh, we utilize the exact (hydrodynamics) Riemann solver, and for self-gravity of the gas envelope we utilize the standard tree-based algorithm. To prevent the loss of the scattering gas mass, the box size is specified as approximately 100 times larger than the initial radius of the binary system, with periodic boundary conditions.

\begin{figure}
\centering
    \includegraphics[width=1\textwidth]{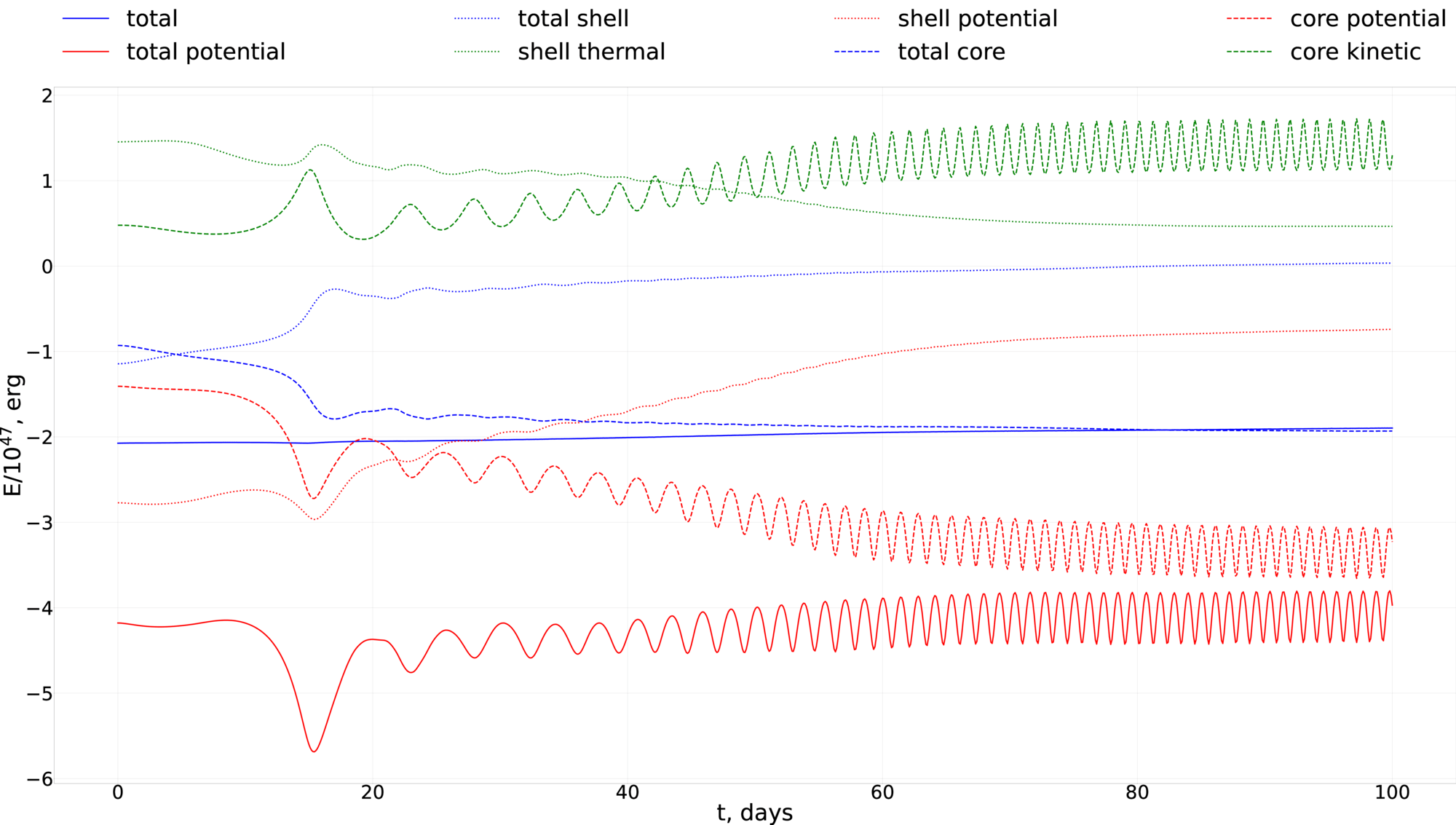}
    \caption{Energy balance for a common envelope phase in Newtonian gravity.}    
    \label{fig:energy_diagrams_Newton}
\end{figure}

Our test calculations show that the accuracy of calculations for RG consisting of $5\times 10^5$ particles is insufficient. On a scale of 100 days, the relative loss of initial total energy, which should be conserved, reaches up to 16 \%. We use the model with $\sim 1.1\times 10^6$ particles because this allows to reduce the energy loss to 6\%. The energy balance for the cores and the shell is depicted on Fig. \ref{fig:energy_diagrams_Newton}. From these dependencies one can see that, in the process of evolution, thermal energy of the shell converts to kinetic energy of the cores.


\subsection{Modified gravity implementation in AREPO}

For our numerical calculations in modified gravity, we implemented  ``modified gravity'' into AREPO code \cite{Springel_2010,Pakmor_2016,Weinberger}. This amendment allows to simulate hydrodynamic processes taking into account the corrections from the GR or theories of modified gravity. Since the AREPO code models both the $N$-bodies problem and hydrodynamic processes based on SPH simulations using the determination of massive particles of various types, we carry out the accounting of corrections to the gravitational field solely through the massive particles, adjusting the gravitational potential that they create, therefore, it being enough to determine the local area in which relativistic effects are significant, while the rest of the global space is modeled within the framework of the Newton's law of universal gravitation. This approach is justified by the fact that deviations from Newtonian gravity contribute only in vicinity of compact and massive gravitational objects. As a result, the corrections due to modification of gravity should be taken into account while modeling gas accretion on compact objects, and while modeling dynamics of material particles near compact objects. 

\begin{figure}
    \centering
        \includegraphics[width=1\textwidth]{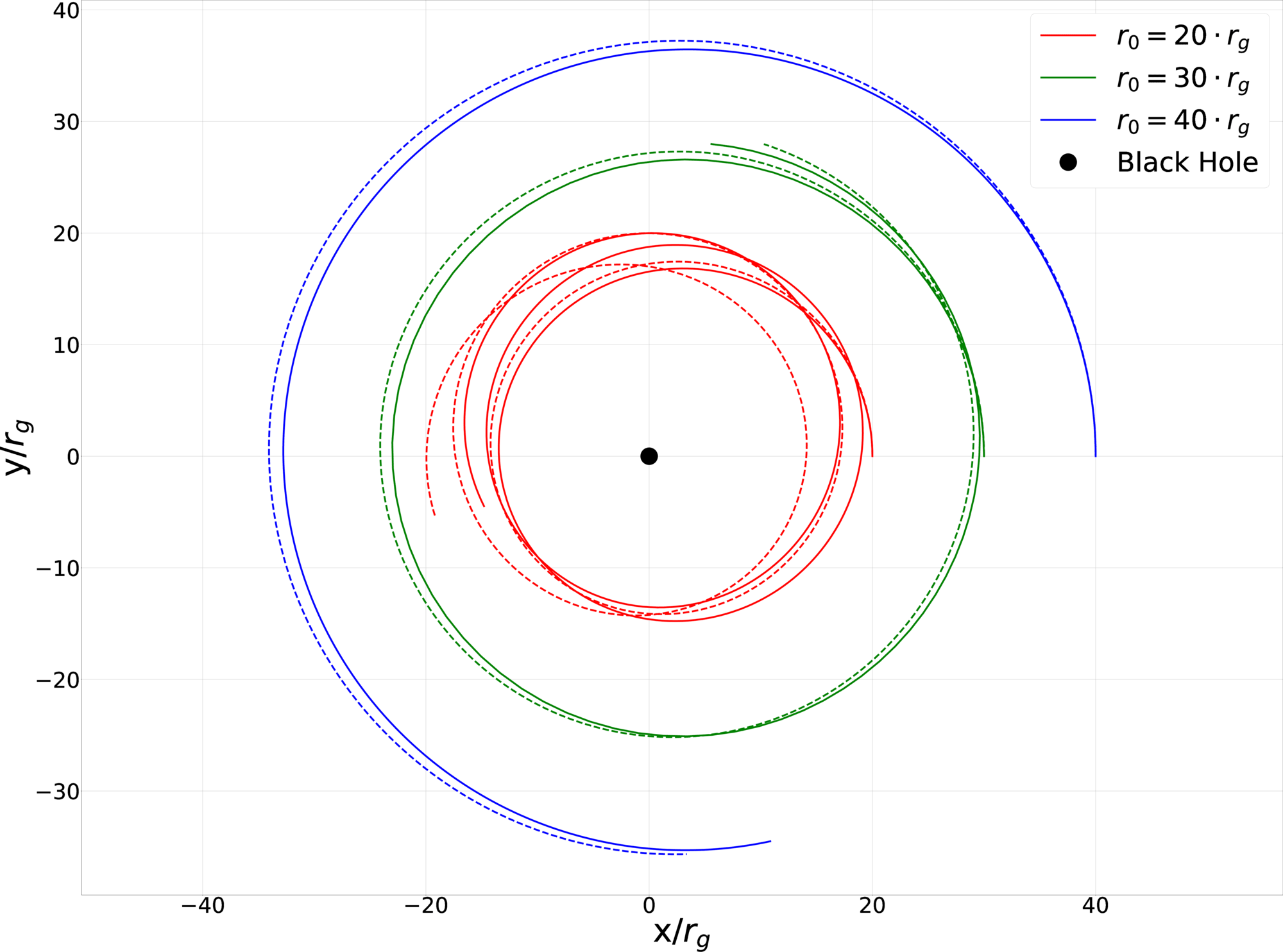}
        \caption{The trajectories of the probe mass moving in the vicinity of the compact point mass (black hole) with corrections from General Relativity implemented into AREPO code (dotted lines)  taken into account, in comparison with the exact solution (solid lines). For the initial conditions we assumed $r_0=20, 30, 40r_g$ and velocity which corresponds to stable circle orbit in Newtonian gravity for each of these radii. The mass of the compact object is 1 $M_{\odot}$ and $r_g = GM_{\odot}/c^2$. \textcolor{blue}{We evolved the system for $1000 r_g /c$}.}
        \label{fig:comparison_GR_Newton_trajectory}
\end{figure}

As a test, we utilize AREPO code to calculate trajectories of the probe mass in vicinity of the point mass (black hole) with corrections from simple Schwarzschild solution for field. The results are compared with exact solutions and shown on the Fig. \ref{fig:comparison_GR_Newton_trajectory}. We see that accuracy of numerical calculations depends on distance between the point mass and the probe mass. For distances around tens and hundreds of gravitational radii, the difference between the exact solution for General Relativity and the solution with corrections is negligible. In our calculations for common envelope phase, the radius of RG with mass $1 M_{\odot}$ is assumed to be $7000$ km and, therefore, for gas particles moving around the RG condition $r>>r_g$ is satisfied. Thus, \textcolor{blue}{we expect} that approximation of pseudo-newtonian potential is valid for the considered task. 

\section{Pseudo-newtonian potential in general relativity and modified gravity}


Typically, in astrophysics and astronomy numerical simulations of various processes are performed  using Newtonian potential. General relativistic equations are complex but some relativistic effects can be reproduced by including the corresponding (pseudo)potential. With this potential we can get the solutions of the hydrodynamical equations.

For example, Paczynski and Wiita (1980) proposed a pseudo-Newtonian potential for Schwarzschild geometry  which can define the properties of the inner disk in equatorial plane around a black hole without using relativistic fluid equations. This potential can be read as
\begin{equation}
\mathcal{U} = -\frac{GM}{r-2r_g},
\end{equation}
where $M$ is the mass of central object and $r_g$ is one half of gravitational radius, $r_g = GM/c^2$. Then, we use a system of units in which $G=c=1$. Considering two first terms in Taylor expansion on $r/r_g<<1$, we obtain
\begin{equation}
\mathcal{U} = -\frac{M}{r} - \frac{2 M r_g}{r^2} = \mathcal{U}_{N} - \frac{2 M r_g}{r^2},
\end{equation}
where $\mathcal{U}_N$ is the usual Newtonian potential. Now consider the simple approach to obtain the (pseudo)potential in a general relativistic case \textcolor{blue}{(see for example \cite{Mukhopadhyay})}. We propose that static spacetime around a compact star possesses spherical symmetry. Therefore, spacetime interval has the form
\begin{equation}
    \label{eq:Spacetime interval}
    ds^2 = g_{tt}dt^2 + g_{rr}dr^2 + g_{\theta\theta}d\theta^2 + g_{\varphi\varphi}d\varphi^2 .
\end{equation}

We consider the test particle with mass $m$ which moves in the equatorial plane, therefore $d\theta = 0, \theta = \frac{\pi}{2}$.  Due to the spherical symmetry, metric functions depend only on radial coordinate $g_{\alpha\beta}=g_{\alpha\beta}(r)$. The Lagrangian density for a particle in the  spacetime at the equatorial plane can be written as 
\begin{equation}
    \label{eq:Proper spacetime interval}
    \mathcal{L} = \frac{m}{2}\left( \frac{ds}{d\tau} \right)^2 = \frac{m}{2}\left(g_{tt}\dot{t}^2 + g_{rr}\dot{r}^2 + g_{\varphi\varphi}\dot{\varphi}^2\right), \qquad \left( \dot{f} = \frac{df}{d\tau} \right),
\end{equation}
where  $\tau$ is the proper time. Let us write the Euler-Lagrange equation $\frac{\partial \mathcal{L}}{\partial q} = \frac{d}{d\tau} \left( \frac{\partial \mathcal{L}}{\partial \dot{q}} \right)$ where $q$ are coordinates ($q = t,r,\theta,\varphi$). 

Note that $t$ or $\varphi$ do not appear in \eqref{eq:Proper spacetime interval}. These variables correspond to two integrals of motion, namely energy and \textcolor{blue}{angular} momentum:
\begin{equation}
	\label{eq:Euler-Lagrange equations}
    -E = \frac{\partial \mathcal{L}}{\partial \dot{t}} = m g_{tt}\dot{t}, \qquad \lambda = \frac{\partial \mathcal{L}}{\partial \dot{\varphi}} = m g_{\varphi\varphi}\dot{\varphi} .
\end{equation}

Let us express $r$ in terms of $\tau$: $r=r\left( \tau \right)$. This can be done directly through \eqref{eq:Euler-Lagrange equations}, or it could also be done differently – through the square of the momentum $\mathbf{p}$ (we use signature $(-,+,+,+)$):
$$
    p^2 = \mathbf{p} \cdot \mathbf{p} = g_{\mu\nu} p^\mu p^\nu = -m^2 c^4
$$
where 
\begin{align*}
    & p_r = \frac{\partial \mathcal{L}}{\partial \dot{r} } = m g_{rr}\dot{r}, & \quad p_\varphi & = \frac{\partial \mathcal{L}}{ \partial \dot{\varphi} } = m g_{\varphi\varphi}\dot{\varphi}, & \quad p_t & = \frac{\partial \mathcal{L}}{\partial \dot{t}} = m g_{tt}\dot{t}, \\
    & p^r = g^{r\alpha}p_{\alpha} = m \dot{r}, & \quad p^\varphi & = g^{\varphi\alpha}p_{\alpha} = m \dot{\varphi}, & \quad p^t & = g^{t\alpha}p_{\alpha} = m \dot{t}.
\end{align*}

Therefore,
$$
    p^2 = g_{tt}(p^t)^2 + g_{rr}(p^r)^2 + g_{\varphi\varphi}(p^\varphi)^2 = g_{tt}\dot{t}^2 + g_{rr}\dot{r}^2 + g_{\varphi\varphi}\dot{\varphi}^2 = -m^2.
$$
Then,
$$
    \dot{r}^2 = - \frac{c^2}{g_{rr}} - \frac{g_{tt}}{g_{rr}}\dot{t}^2 - \frac{g_{\varphi\varphi}}{g_{rr}}\dot{\varphi}^2.
$$
Further on we express $\dot{t}$ and $\dot{\varphi}$ in terms of \textcolor{blue}{$\tilde{E}=E/m$ and $\tilde{\lambda}=\lambda/m$}:
$$
    \dot{t} = -\frac{\tilde{E}}{g_{tt}},\quad \dot{\varphi} = \frac{\tilde{\lambda}}{ g_{\varphi\varphi}}.
$$
Then,
$$
    \dot{r}^2 = - \frac{c^2}{g_{rr}} - \frac{g_{tt}}{g_{rr}}\left(\frac{\tilde{E}}{g_{tt}}\right)^2 - \frac{g_{\varphi\varphi}}{g_{rr}}\left(\frac{\tilde{\lambda}}{g_{\varphi\varphi}}\right)^2 = - \frac{c^2}{g_{rr}} - \frac{\tilde{E}^2}{g_{tt}g_{rr}} - \frac{\tilde{\lambda}^2}{g_{\varphi\varphi}g_{rr}} = \Phi(r,\tilde{E},\tilde{\lambda}).
$$
If the orbits are stable circles, then $\dot{r} = 0$, and 
\begin{equation}
    \label{eq:system on Phi}
    \Phi(r,E,\lambda) = 0,\quad \frac{d\Phi}{dr} = 0.
\end{equation}

From \eqref{eq:system on Phi} one can find $E$ and $\lambda$. In the case of spherical symmetric metric, one can choose the coordinates such that the angular part of the metric can be written as $r^2 d\Omega^2$. So, we can assume that $g_{\varphi\varphi}=r^2$, thus
\begin{equation}\label{sch-metr}
    ds^2 = g_{tt}dt^2 + g_{rr}dr^2 + r^2 d\varphi^2.
\end{equation}
Therefore, from \eqref{eq:system on Phi} we obtain
\begin{equation}
    \label{eq:relation between E and lambda}
    \frac{\tilde{E}^2}{g_{tt}} + \frac{\tilde{\lambda}^2}{r^2} = -1
\end{equation}
We differentiate \eqref{eq:relation between E and lambda} by $r$
$$
    \frac{d\Phi}{dr}\equiv\frac{1}{g^2_{rr}}g_{rr,r} + \frac{\tilde{E}^2}{g^2_{rr}g^2_{tt}}(g_{rr}g_{tt,r} + g_{rr,r}g_{tt}) + \frac{\tilde{\lambda}^2}{g^2_{rr}r^4}(2rg_{rr} + g_{rr,r}r^2) = 0.
$$
The obtained  equation can be simplified:
\begin{equation}
    \label{eq:very long equation}
    g_{rr,r} + \frac{\tilde{E}^2}{g^2_{tt}}(g_{rr}g_{tt,r} + g_{rr,r}g_{tt}) + \frac{\tilde{\lambda}^2}{r^4}(2rg_{rr}+g_{rr,r}r^2) = 0.
\end{equation}

We solve the system \eqref{eq:relation between E and lambda}, 
\eqref{eq:very long equation} with respect to $E$ and $\lambda$:
$$
{\tilde{\lambda}^2} = \frac{r^3g_{tt,r}}{2g_{tt} - rg_{tt,r}}, 
$$
$$
{\tilde{E}^2} = - \frac{2g^2_{tt}}{2g_{tt} - rg_{tt,r}}.
$$
Let us introduce
$$
    \lambda^2_K = \frac{\tilde{\lambda}^2}{\tilde{E}^2} = -\frac{r^3g_{tt,r}}{2g^2_{tt}}.
$$
Notice that $-\frac{\lambda^2_K}{r^3}$ is the centrifugal acceleration of a particle in the gravitational field. Thus, we obtain 
$$
    F_{c} = \frac{g_{tt,r}}{2g^2_{tt}}.
$$
For example, in the case of the Schwarzschild metric we have
$$
    g_{tt} = - \left( 1-\frac{2 r_g}{r} \right),
$$
Then $g_{tt,r} = -  \frac{2r_g}{r^2}$, and
$$
    F_{c} = - \frac{2 r_g}{2r^2(1-{2r_g}/r)^2} = -\frac{M}{(r-2r_g)^2}.
$$
Therefore, the potential is
$$
    \mathcal{U} = -\int F_{c}dr = - \frac{M}{r - 2r_g}.
$$
In general, the (pseudo)potential can be expressed explicitly up to a constant
\begin{equation}
    \mathcal{U} = - \int F_{c} dr = -\int \frac{g_{tt,r}}{2g^2_{tt}} dr = -\frac{1}{2} \int g_{tt}^{-2} dg_{tt} = \frac{1}{2g_{tt}} + C .
\end{equation}

For gravitational field equation, in the case of some $F(R)$ gravity, where $R$ is scalar curvature, we start from gravitational field action:
\begin{equation}\label{action}
S_g=\frac{1}{16\pi}\int d^4x \sqrt{-g}F(R),
\end{equation}
For solution describing the spherically symmetric field around the star (in a space without matter), one can assume the metric in form (\ref{sch-metr}). Varying the action with respect to the  metric tensor elements gives the following equation for metric functions:
\begin{equation}\label{field}
F'(R)G_{\mu \nu }-\frac{1}{2}(F(R)-F'(R)R)g_{\mu \nu }-(\nabla
_{\mu }\nabla _{\nu }-g_{\mu \nu }\Box )F'(R)=0.
\end{equation}
Here $G_{\mu\nu}=R_{\mu\nu}-\frac{1}{2}Rg_{\mu\nu}$ is the
Einstein tensor, $F'(R)=dF(R)/dR$. For some $F(R)$ function $g_{tt}$ can be written in an explicit form. We consider the model with the following $g_{tt}$: 
$$
g_{tt} = -1+\frac{2M}{r} - \frac{\beta}{3} \left(1 - e^{-\gamma r^2} \right).
$$
Where $\beta$ is dimensionless constant and $\gamma$ is constant with dimension of inverse square root of lenght. The corresponding (pseudo)potential is equal to
\begin{equation}\label{U_0}
    \mathcal{U} = \frac{3}{2 \beta + 6} - \frac{3r e^{\gamma r^2}}{ \left( 2\beta r + 6r -12M \right) e^{\gamma r^2} - 2r\beta}, \quad \gamma > 0.
\end{equation}
For $r>>2M$ $g_{tt}\rightarrow -1 - \beta/3$ and therefore we have flat spacetime on large distances. For $r<<\gamma^{-1/2}$ we have usual Schwarzschild geometry around compact star. Therefore possible effects from modification of gravity can take place on intermediate distances from compact star.

\section{Results}

\begin{figure}
\centering
    \includegraphics[width=0.9\textwidth]{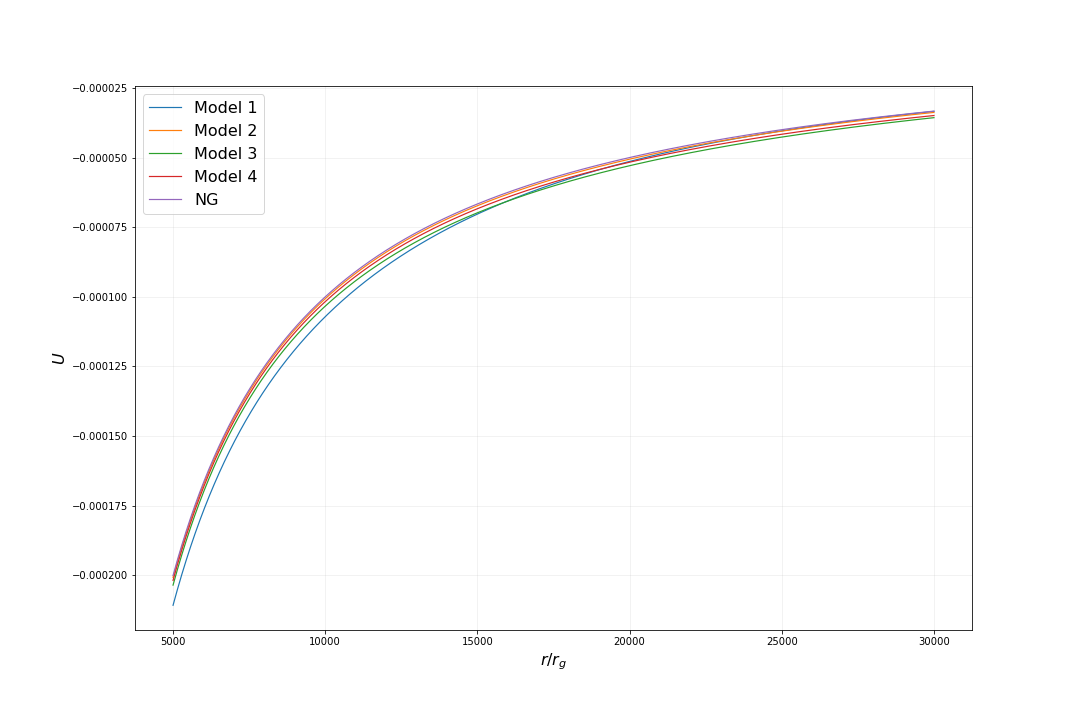} 
    \caption{\textcolor{blue}{The (pseudo)potential $\mathcal{U}$ (in units of $c^2$) as function of dimensionless radial coordinate $r/r_g$ from surface of WD ($\sim 5000 r_g$) up to $\sim 5$ radii of it.}  } 
    \label{fig:eff_potential}
\end{figure}

We considered the evolution of the binary system described above in Newtonian gravity and modified gravity. For our model of modified gravity we redefine parameter $\gamma$ as 
$$
\gamma = 1/L_{0}^2,
$$
where $L_{0}$ is some scale. Parameter $\beta$ can be written as follows:
$$
\beta = \beta_{0} r_g / L_0.
$$
The following parameters have been considered ($L_0$ is given in $10^6$ km):  $L_0 = 0.02$, $\beta_0 = 1$ (Model 1), $L_0 = 0.07$, $\beta_0=0.2$ (Model 2), $L_0 = 0.07$, $\beta_0=1$ (Model 3), $L_0=0.14$, $\beta_0=1$ (Model 4). Therefore we consider scales $L_0$ which are comparable with radius of RG ($r_{s}=0.007$ in given units). Parameter $\beta$ is very small, because $r_g=1.47$ km for RG with $M=M_{\odot}$ and for considered parameters $\beta<10^{-4}$. \textcolor{blue}{We depicted corresponding (pseudo)potentials for these parameters on Fig. \ref{fig:eff_potential}.}

\begin{figure}
\centering
    \includegraphics[width=0.47\textwidth]{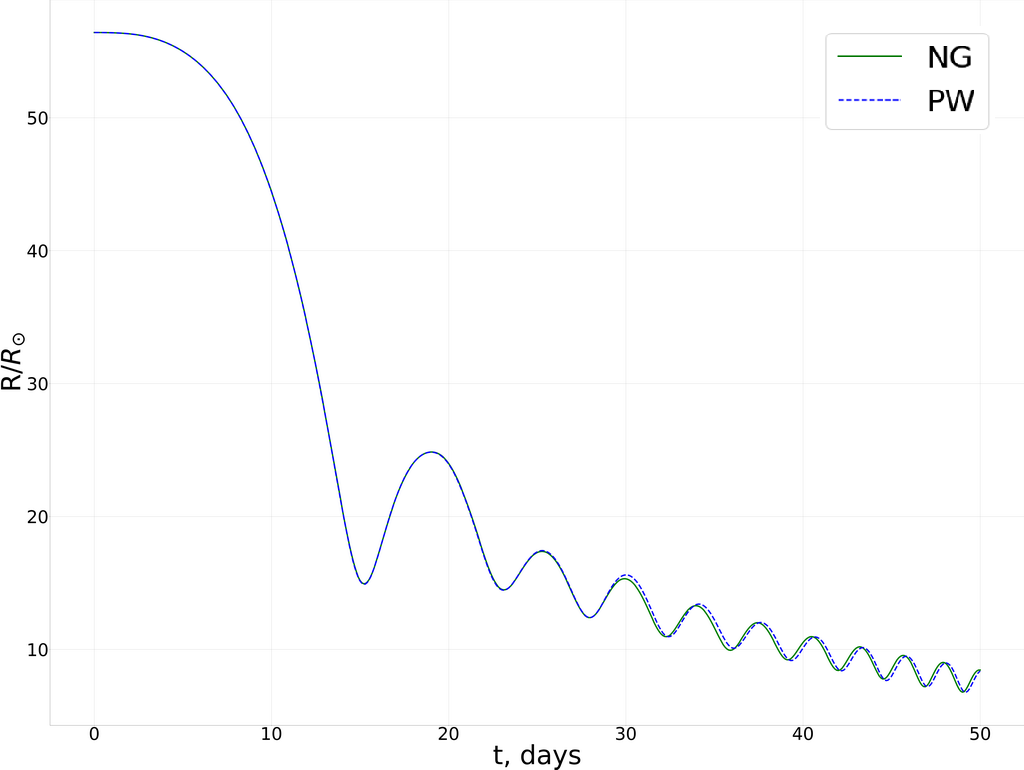} \includegraphics[width=0.47\textwidth]{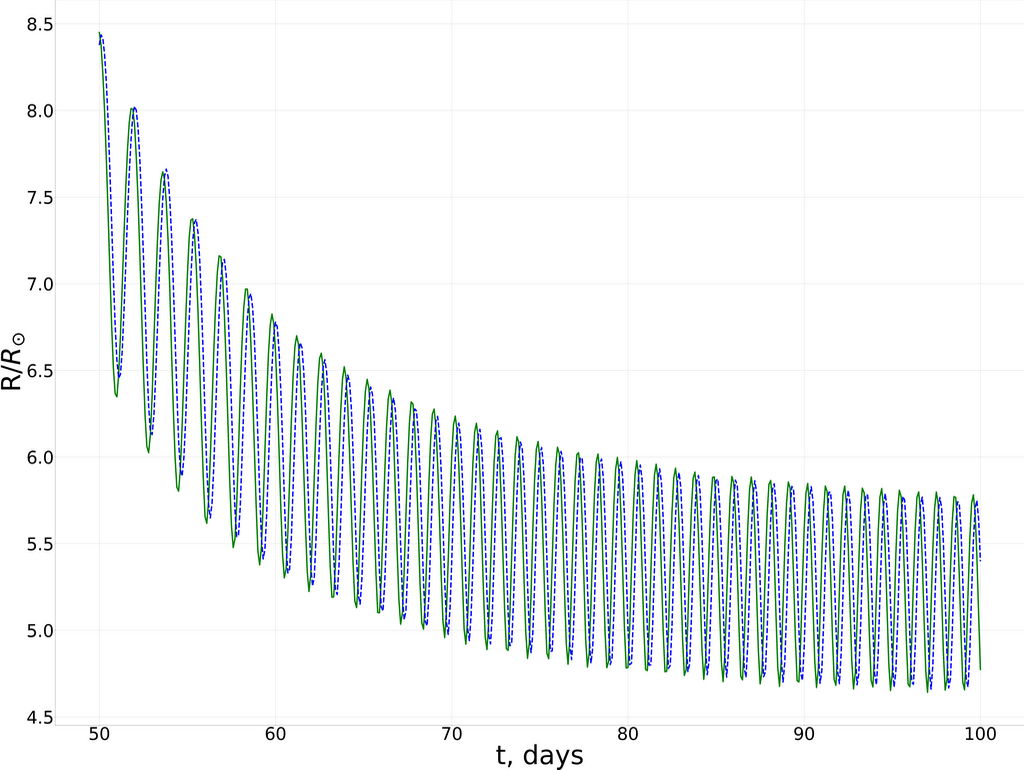}
    \caption{Distance between RG core and WD in solar radii as function of time \textcolor{blue}{for the case of Newtonian gravity (NG) and Paczynski–Wiita potential (PW)}  in interval between days 0 and 50 (left panel) and between days 50 and 100.}    
    \label{fig:time_distance_and_orbits_graph_Newton}
\end{figure}

\begin{figure}
    \centering
    \includegraphics[width=0.45\linewidth]{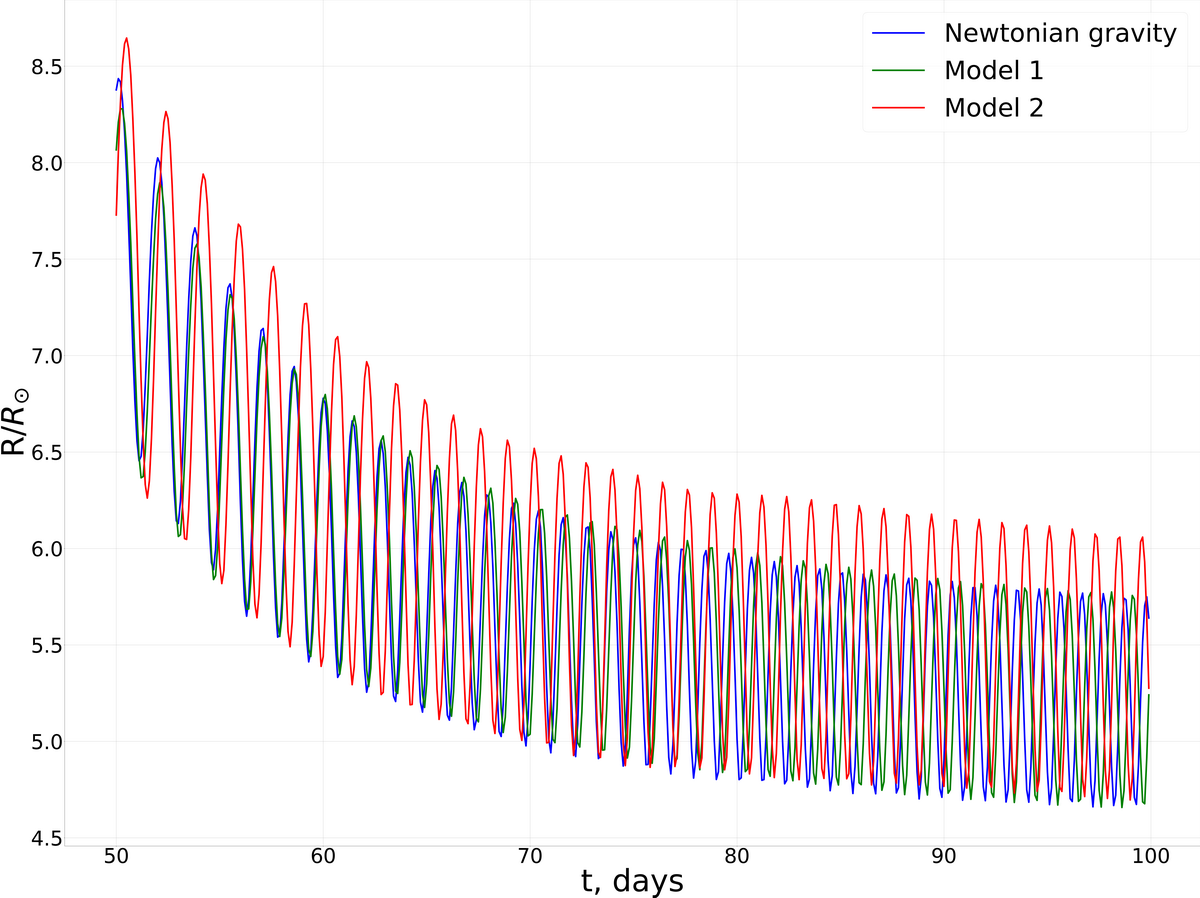}
    \includegraphics[width=0.45\linewidth]{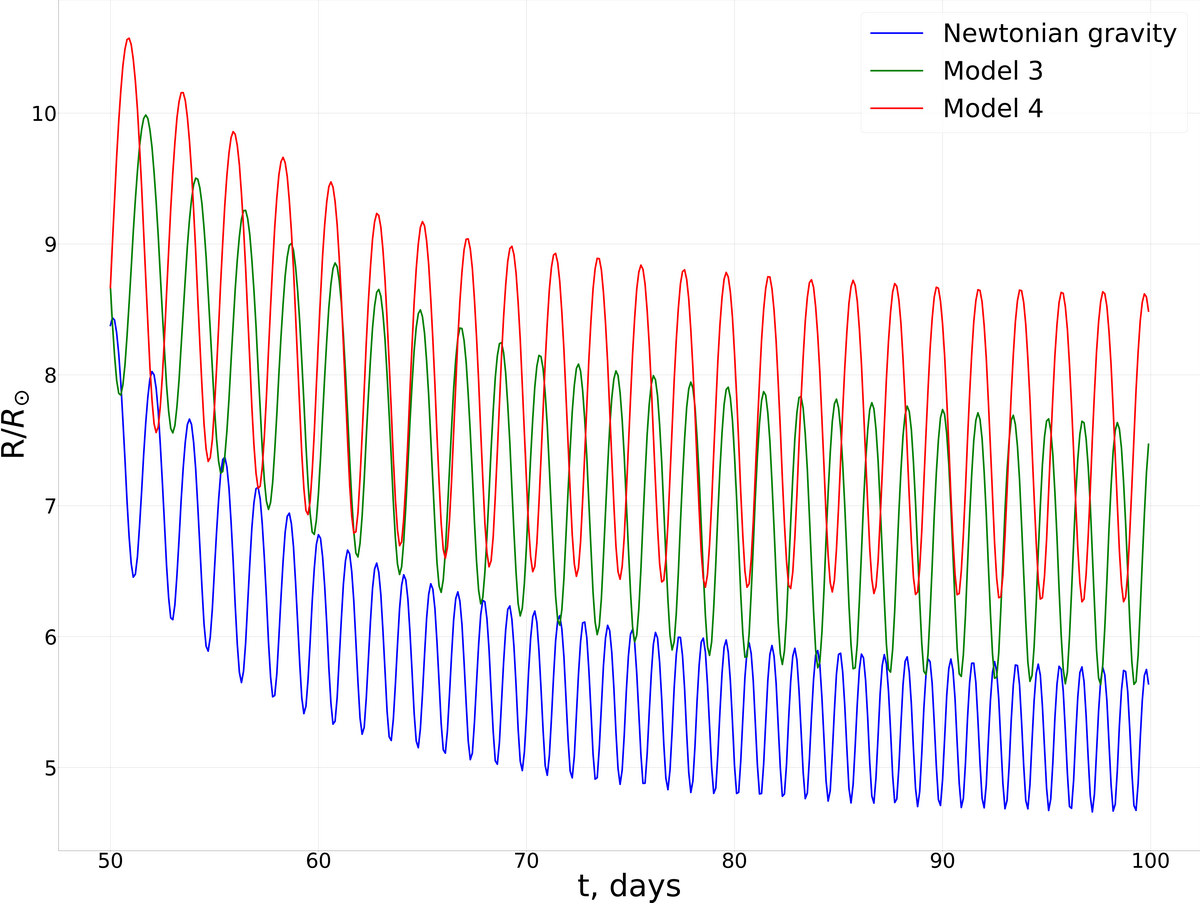}
    \caption{The same as on Fig. \ref{fig:time_distance_and_orbits_graph_Newton} but for models of gravity \ref{U_0} with various parameters $\beta$ and $\gamma$ in interval between days 50 and 100.}
    \label{fig_distance}
\end{figure}

We simulated the movement of system components over the course of 100 days. As expected, the first stage, namely rapid \textcolor{blue}{inspiral} (up to $t\approx ~20$ days) is very similar for Newtonian and modified gravity. To illustrate that  on Fig. \ref{fig:time_distance_and_orbits_graph_Newton} we show the dependence of distance between the RG core and the WD on time for Newtonian gravity and corrections due to the Schwarzschild solution \textcolor{blue}{for gravitational field around WD}. This difference is negligible. Then, we can see the difference between Newtonian gravity and modified gravity (see Fig. \ref{fig_distance}). This difference reflects some repulsion effect depending on the additional terms in (pseudo)potential. After several orbits, the distance between the RG core and the WD decreases much slower in comparison to the beginning of the simulation. For modified gravity, at certain parameters the mean distance between the components at the end of the simulation is larger than it would have been for usual gravity (see Table 1). The major axis and eccentricity of orbit depend strongly on parameter $L_0$ for fixed $\beta_0$. The corresponding number of orbits decreases. While for Newtonian gravity the components revolved around each other approximately 50 times, for modified gravity with $\beta_0=1$ these numbers are $30 - 35$ revolutions for 100 days in a the case of $L_0 = {0.07}$ and $L_0 = 0.14$.

\begin{table}
\label{Table1}
\begin{centering}
\begin{tabular}{|c|c|c|c|c|c|}
  \hline
   & $L_0$ & $\beta_0$ & $a$ & $e$ & $N$ \\
  Model & ($10^6$ km) &  & ($R_{\odot}$) & &   \\
  \hline
  NG & 0 & 0 & 5.2 & 0.095 & 50 \\
  1 & 0.02 & 1 & 5.2 & 0.095 & 50 \\
  2 & 0.07 & 0.2 & 6.7 & 0.15 & 48 \\
  3 & 0.07 & 1 & 7.1 & 0.15 & 36  \\
  4 & 0.14 & 1 & 7.5 & 0.16 & 33  \\  
  \hline
\end{tabular}
\caption{\textcolor{blue}{The list of models with parameters and corresponding data such as major axis $a$, eccentricity $e$ and number of revolutions $N$ for 100 days. For comparison results for Newonian gravity (NG) are also given.}}
\end{centering}
\end{table}

\begin{figure}
    \centering
    \includegraphics[width=0.9\linewidth]{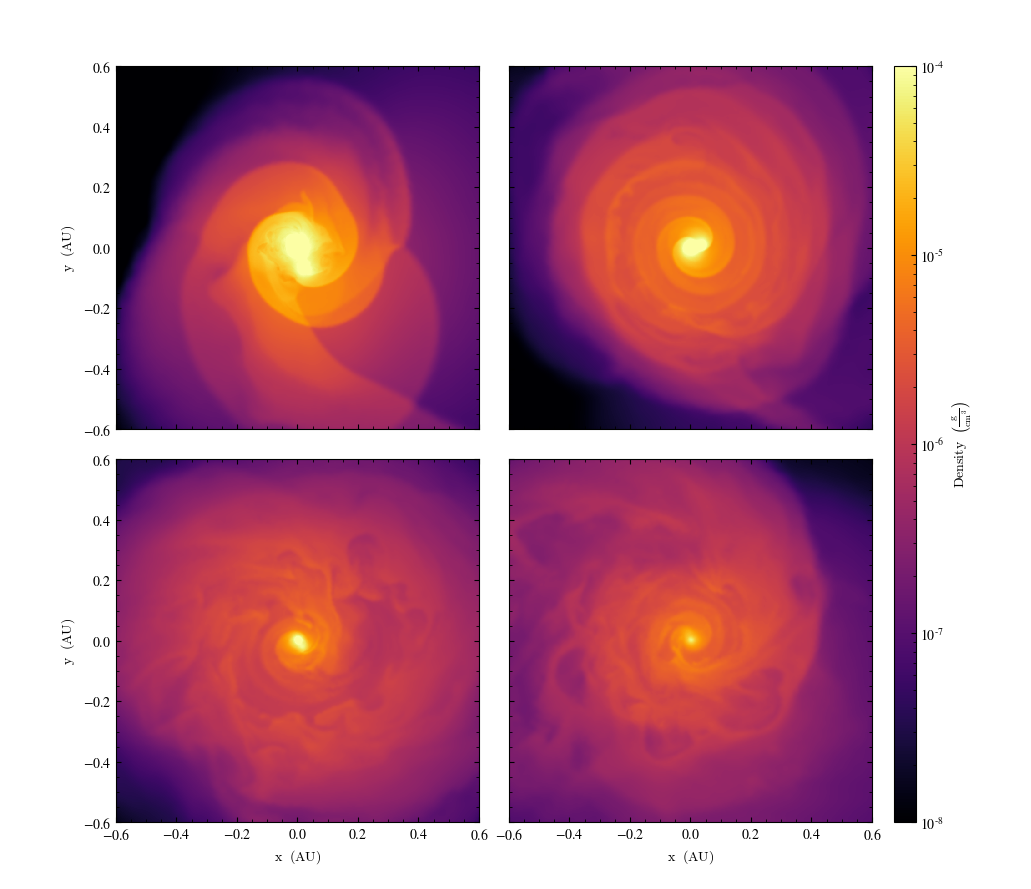}
     \caption{Density slices in $(x,y)$-plane ($z = 0$) for different moments in time in a case of Newtonian gravity: $t=25$ days (left upper panel), $t=50$ days (right upper panel), $t=75$ days (left down panel), $t=100$ days (right down panel). Plots are centered on the point in which gas density is maximal.}
    \label{fig_density_slice_NG}
\end{figure}

\begin{figure}
    \centering
    \includegraphics[width=0.9\linewidth]{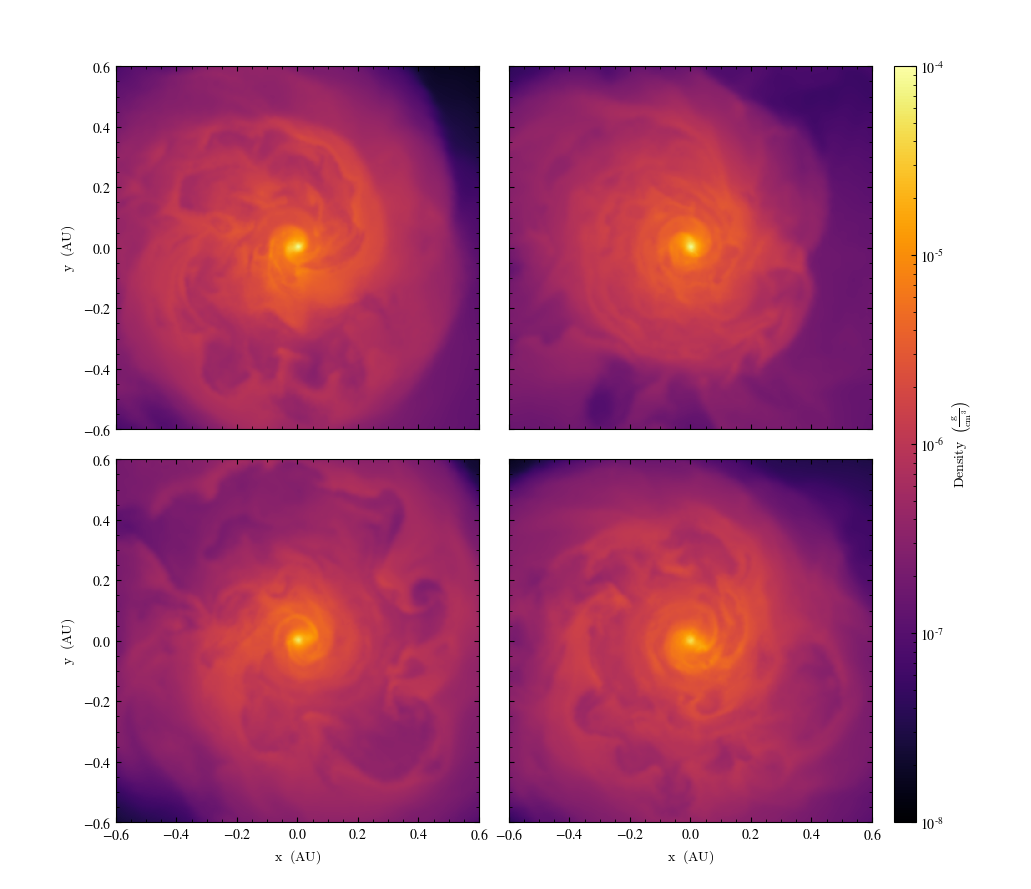}
    \caption{Density slices in $(x,y)$-plane ($z = 0$) for the last moment of simulation $t=100$ days in a case of modified gravity with various parameters for Model 1 (left upper panel), Model 2 (right upper panel), Model 3 (left down panel) and Model 4 (right down panel).} 
    \label{fig_density_slice_MG}
\end{figure}

The evolution of RG envelope and accretion stream were also investigated. We illustrated this process using the slices of gas density in x–y plane (see Figs. \ref{fig_density_slice_NG} and \ref{fig_density_slice_MG}). On the initial stage the picture is similar for Newtonian and modified gravity, which can be described as follows. The accreting gas during the first revolution creates spiral shock waves and the orbit shrinks rapidly. Then, a layered structure appears due to the shock waves. On the late stage the shear flows between close shock waves lead to the emergence of Kelvin–Helmholtz instabilities. 

It is important to note that the accretion stream is asymmetric due to tidal forces, and center of mass of the RG core and the RG shifts from the initial position (for the considered modification of gravity this shifting is smaller in comparison to Newtonian gravity). The envelope is ejected in the opposite direction.

On Fig. \ref{fig_density_slice_MG} we demonstrate the structure of the envelope in the plane of rotation at the end of the simulation for modified gravity after 100 days. Having compared this plot with the last snapshot of Fig. \ref{fig_density_slice_NG} one can conclude, that Kelvin–Helmholtz instabilities are intensifying for modified gravity. On such a scale the layered structure becomes weak and the gas flow is governed by these instabilities.


\begin{figure}
    \centering
    \includegraphics[width=0.8\linewidth]{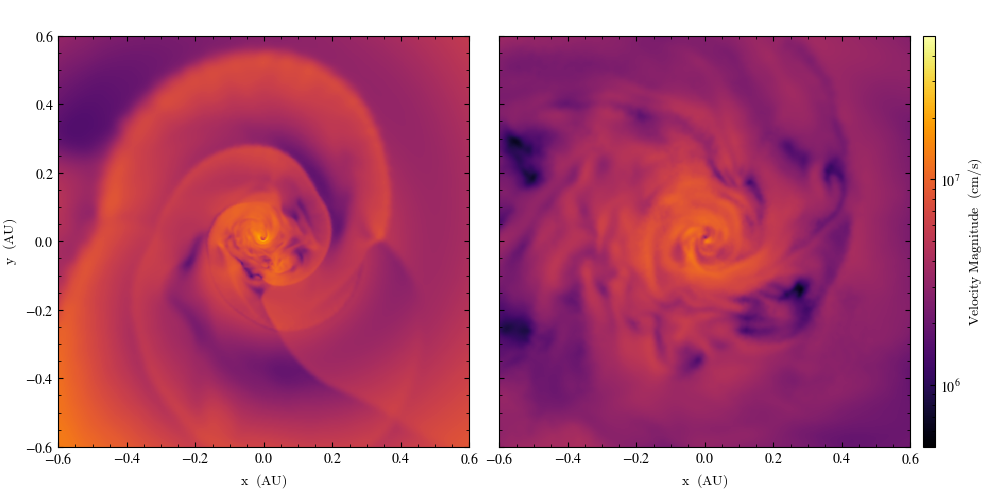}
    \caption{Velocity slices in $(x,y)$ plane ($z=0$) in a case of Newtonian gravity: $t=25$ days (left panel), $t=100$ days (right panel). Plots are centered on the point in which gas density is maximal.}
    \label{fig_velocity_slice_NG}
\end{figure}

\begin{figure}
    \centering
    \includegraphics[width=0.9\linewidth]{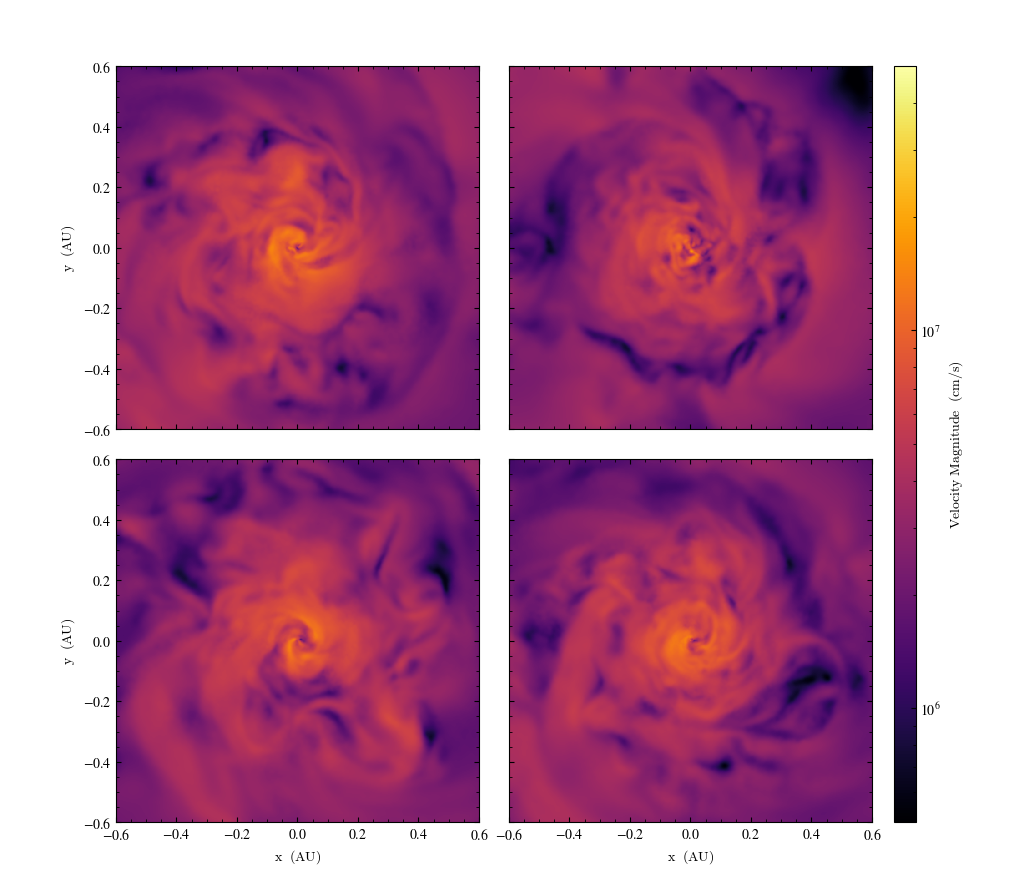}
    \caption{Velocity slices in $(x,y)$ plane ($z=0$) for the last moment of simulation $t=100$ days in a case of modified gravity with various parameters for Model 1 (left upper panel), Model 2 (right upper panel), Model 3 (left down panel) and Model 4 (right down panel).}
    \label{fig_velocity_slice_MG}
\end{figure}


\begin{figure}
    \centering
    \includegraphics[width=0.8\linewidth]{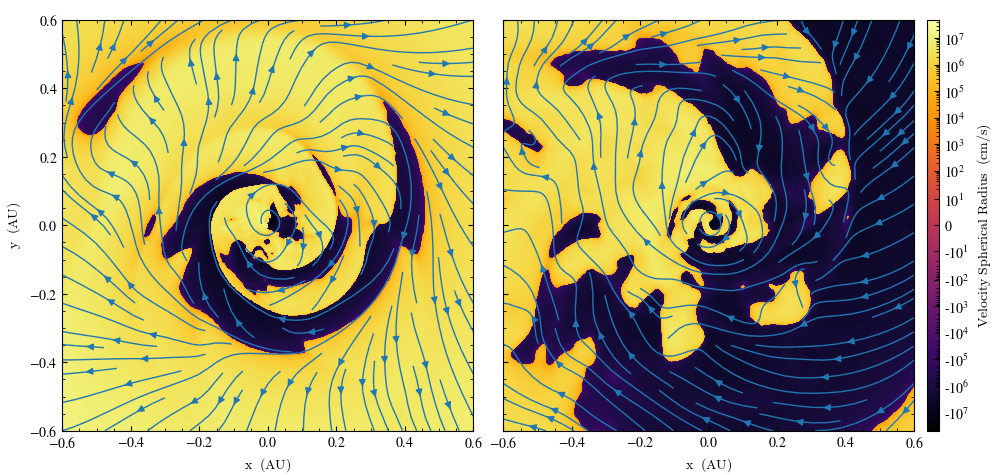}
    \caption{The same as on Fig. \ref{fig_velocity_slice_NG} but for radial velocity in $(x,y)$-plane ($z = 0$) with corresponding streamlines.} 
    \label{fig_velocity_radial_NG}
\end{figure}

\begin{figure}
    \centering
    \includegraphics[width=0.9\linewidth]{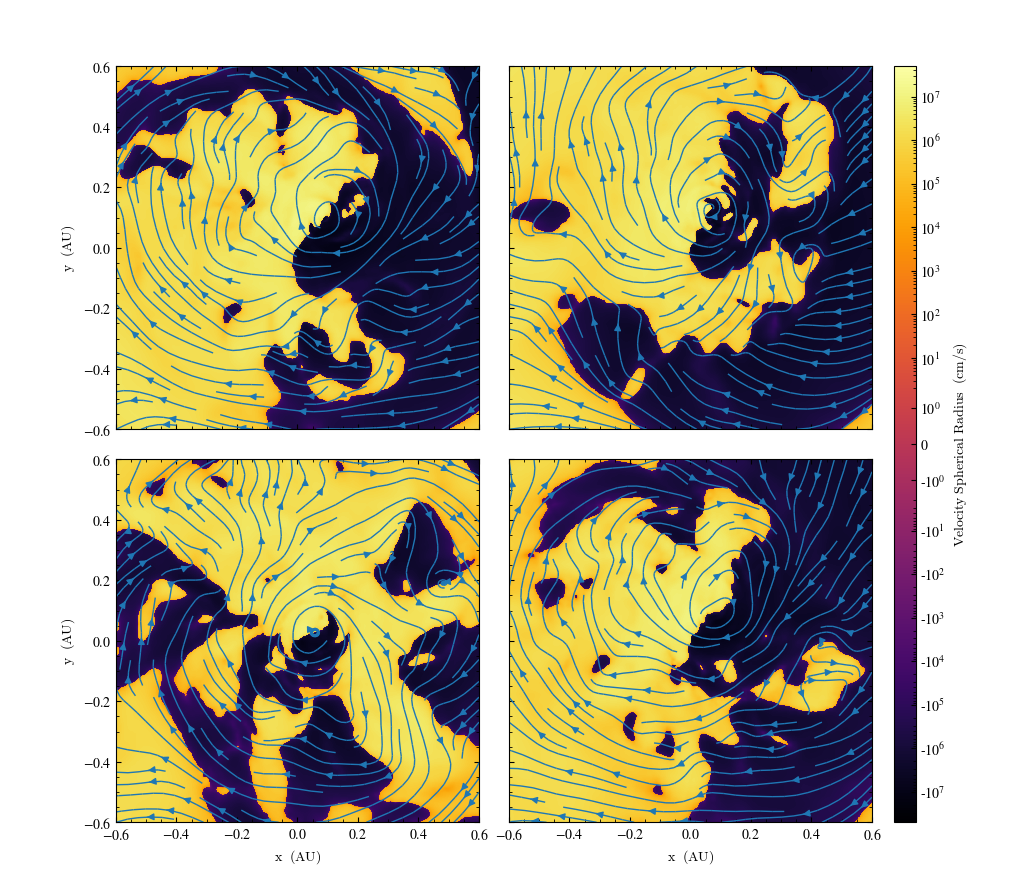}
    \caption{The same as on Fig. \ref{fig_velocity_slice_MG} but for radial velocity in $(x,y)$-plane ($z = 0$) with corresponding streamlines.} 
    \label{fig_velocity_radial_MG}
\end{figure}

As for density, we depicted the field of velocity for four moments in time in a case of Newtonian gravity (Fig. \ref{fig_velocity_slice_NG}) and last snapshots for modified gravity with various parameters (Fig. \ref{fig_velocity_slice_MG}). Again, we see some interesting features in a case of modified gravity. For some parameters instabilities grow faster than for Newtonian gravity. Analysis of radial velocity in the orbital plane shows that some areas with center-bound inflow streams emerge in the process of evolution (Fig. \ref{fig_velocity_radial_NG}). For modified gravity at some parameters (for example, down left panel on Fig. \ref{fig_velocity_radial_MG}) this process is slower and after 100 days an outflow stream from the center dominates. This can be explained by a presence of a strong rebound after the initial sharp shrinking of the distance between the RG core and the RG. 

\begin{figure}
    \centering
    \includegraphics[width=0.8\linewidth]{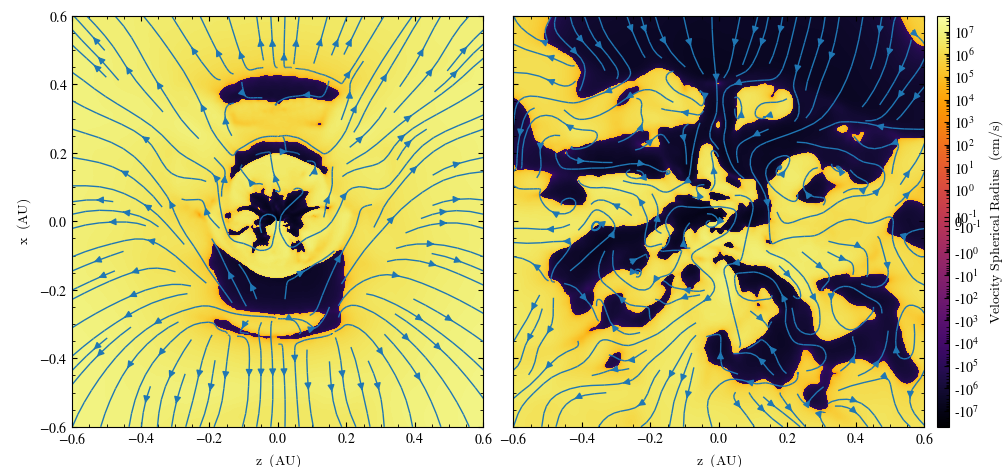}
    \caption{Radial velocity slices in $(x,z)$-plane ($y = 0$) in a case of Newtonian gravity: $t=25$ days (left panel), $t=100$ days (right panel).} 
    \label{fig_velocity_slices_XZ_GR}
\end{figure}

\begin{figure}
    \centering
    \includegraphics[width=0.9\linewidth]{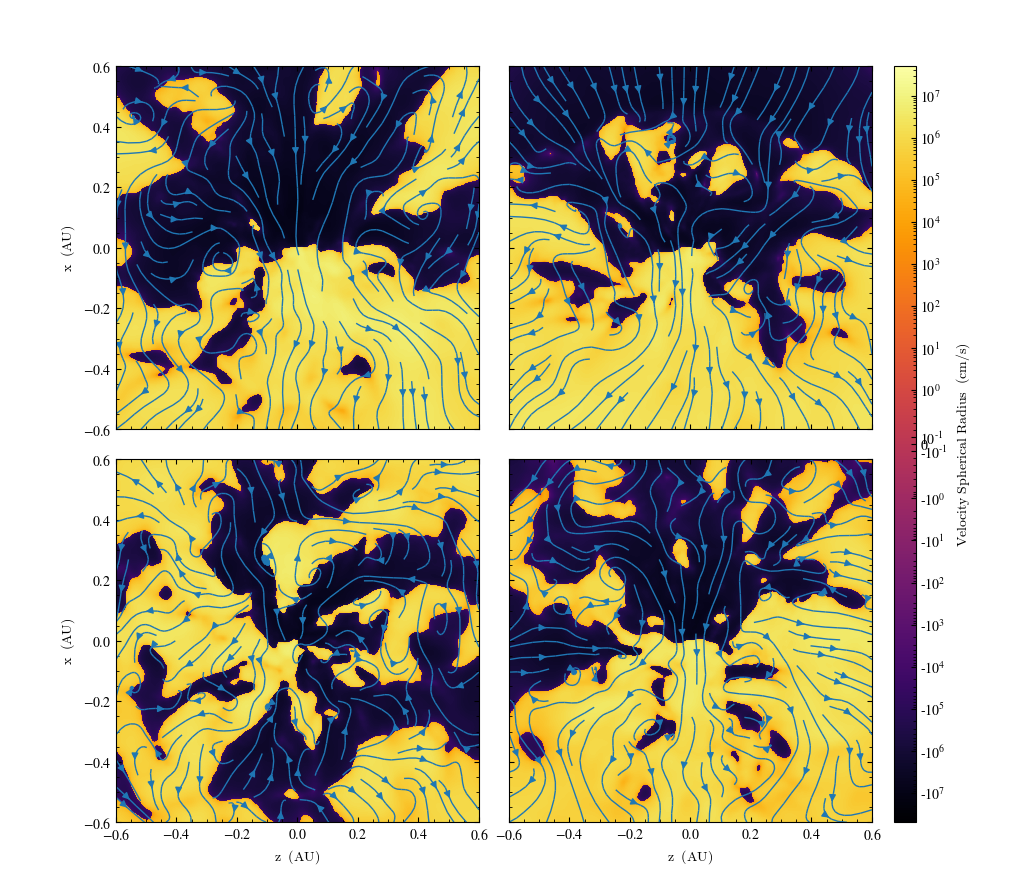}
    \caption{Radial velocity slices in $(x,z)$-plane ($y = 0$) for the last moment of simulation $t=100$ days in a case of modified gravity with various parameters for Model 1 (left upper panel), Model 2 (right upper panel), Model 3 (left down panel) and Model 4 (right down panel).} 
    \label{fig_density_velocity_XZ_MG}
\end{figure}

We also considered the structure of the stream in the direction perpendicular to the orbital plane (Fig. \ref{fig_velocity_slices_XZ_GR}). The center-bound stream of mass is concentrated mostly in area $x>0$ for Newtonian gravity. For some parameters in modified gravity the picture is more complex. In every case we have whirls corresponding to the instabilities in the flow. 

This complex structure makes it
difficult to predict the further evolution of energy transfer in
this plane, although the velocity is
rather small in the region of the instability. Especially in the
inner part, adjacent layers can be found with differing
velocities, resulting in shear flows.
    
\section{Conclusion}

We investigated the structure and the evolution of common envelope forming during the inspiral of components in the binary system with using moving-mesh AREPO code. Our numerical simulations were performed for a case of Newtonian gravity and simple model of modified gravity with exponential term on scalar curvature. The main purpose of our investigation was to find possible imprints of modified gravity in accretion flow. 

\textcolor{blue}{For modified gravity as in a case of Newtonian gravity}, we see that shock waves emerge during the first revolution and the orbit shrinks rapidly. Also, the layered structure appears due to spiral shock waves. Then, we observed    some features in a case of modified gravity. First, repulsion effect from additional terms in (pseudo)potential takes place. Second, the companions of the system approach more slowly, the major axis of the orbit is larger and the orbit is more elongated. Due to this, the number of revolutions for the same time period as for Newtonian gravity decreases. 

However, it is interesting note that large-scale instabilities due to shear between close layers for some parameters of modified gravity may appear more clearly in comparison to Newtonian gravity. Although the considered model of modified gravity looks hypothetical, our calculations may be of interest because we showed that sufficiently small deviations from Newtonian gravity can lead to large consequences for further evolution of closed binary and envelope, because instabilities in hydrodynamical flow lead to turbulence in the future.

\textcolor{blue}{Of course the very difficult question arises. Given how complicated and poorly understood the common envelope
phase is, what missing astrophysical modeling (magnetic field,
etc.) could mimic the effects of modified gravity found? In principle we cannot exclude the possibility of influence of various factors on features of hydrodynamical flow. Our main goal was to investigate the long-time consequences of possible declinations from general relativity in description of gravity (in frames of simple model and fixed parameters such as masses of stars, initial distance between components, etc.)}

As the next step of our work, we plan to explore the further evolution of closed binary system in a case of modified gravity and consider systems with various parameters, such as mass and radius of giant, and consider another model of modified gravity.

\acknowledgments

This work is supported by  Program  ``Priority 2030'',  № 421-L-23  (Immanuel Kant Baltic Federal University, Russia). The authors wish to express gratitude to Sergey Borohov, Lev Oganisyan, Pavel Vasilev, Maksim Tsarkov and Vladimir Bashkirov for installing AREPO code, for providing the necessary technical support for working with the code, as well as for supplying the servers for calculations. A.A. thanks to Rainer Weinberger for answering some questions concerning AREPO code. 

\paragraph{Software:} AREPO code \cite{Weinberger}, yt project \cite{Turk_2011}, matplotlib \cite{Hunter}. 

\paragraph{Declaration of competing interest.} The authors declare that they have no known competing financial
interests or personal relationships that could have appeared
to influence the work reported in this paper.

\paragraph{Data availability.} No new data were generated or analysed
in support of this research.




\bibliographystyle{JHEP}
\bibliography{bibliography}

\providecommand{\href}[2]{#2}\begingroup\raggedright\begin{thebibliography}{10}

\bibitem{Ivanova_2013}
N.~{Ivanova}, S.~Justham, X.~Chen, O.~De~Marco, C.L.~Fryer, E.~Gaburov et~al., \emph{Common envelope evolution: where we stand and how we can move forward}, \href{https://doi.org/10.1007/s00159-013-0059-2}{\emph{Astron. Astrophys. Rev.} {\bfseries 21} (2013) 59}.

\bibitem{Iaconi_2019}
R.~{Iaconi} and O.~{De Marco}, \emph{{Speaking with one voice: simulations and observations discuss the common envelope $\alpha$ parameter}}, \href{https://doi.org/10.1093/mnras/stz2756}{\emph{MNRAS} {\bfseries 490} (2019) 2550}.

\bibitem{Ricker_2012}
P.~{Ricker} and R.~{Taam}, \emph{An amr study of the common-envelope phase of binary evolution}, \href{https://doi.org/10.1088/0004-637X/746/1/74}{\emph{Astrophys. J.} {\bfseries 746} (2012) 74}.

\bibitem{Passy_2012}
J.-C.~{Passy}, O.~{De Marco}, C.~{Fryer}, F.~{Herwig}, S.~{Diehl}, J.~{Oishi} et~al., \emph{Simulating the common envelope phase of a red giant using smoothed-particle hydrodynamics and uniform-grid codes}, \href{https://doi.org/10.1088/0004-637X/744/1/52}{\emph{Astrophys. J.} {\bfseries 744} (2011) 52}.

\bibitem{Ohlmann_2016}
S.~{Ohlmann}, F.~{R{\"o}pke}, R.~{Pakmor} and V.~{Springel}, \emph{{Hydrodynamic Moving-mesh Simulations of the Common Envelope Phase in Binary Stellar Systems}}, \href{https://doi.org/10.3847/2041-8205/816/1/L9}{\emph{Astrophys. J.} {\bfseries 816} (2016) L9}.

\bibitem{Sand_2020}
C.~{Sand}, S.~{Ohlmann}, F.~{Schneider}, R.~{Pakmor} and F.~{Röpke}, \emph{Common-envelope evolution with an asymptotic giant branch star}, \href{https://doi.org/10.1051/0004-6361/202038992}{\emph{Astron. Astrophys.} {\bfseries 644} (2020) A60}.

\bibitem{Riess-1}
A.~{Riess}, {Filippenko, A.V.}, {Challis, P.}, {Clocchiatti, A.}, {Diercks, A.}, {Garnavich, P.M.} et~al., \emph{Observational evidence from supernovae for an accelerating universe and a cosmological constant}, \href{https://doi.org/10.1086/300499}{\emph{Astron. J.} {\bfseries 116} (1998) 1009}.

\bibitem{Perlmutter}
S.~{Perlmutter}, {Aldering, G.}, {Goldhaber, G.}, {Knop, R.~A.}, {Nugent, P.}, {Castro, P.~G.} et~al., \emph{Measurements of {\ensuremath{\Omega}} and {\ensuremath{\Lambda}} from 42 high-redshift supernovae}, \href{https://doi.org/10.1086/307221}{\emph{Astrophys. J.} {\bfseries 517} (1999) 565}.

\bibitem{Riess-2}
A.~{Riess}, {Strolger, L.-G.}, {Tonry, J.}, {Casertano, S.}, {Ferguson, H.C.}, {Mobasher, B.} et~al., \emph{Type ia supernova discoveries at z > 1 from the hubble space telescope: Evidence for past deceleration and constraints on dark energy evolution}, \href{https://doi.org/10.1086/383612}{\emph{Astrophys. J.} {\bfseries 607} (2004) 665}.

\bibitem{Odintsov1}
S.~{Nojiri} and S.~{Odintsov}, \emph{Unifying inflation with {\ensuremath{\Lambda}}cdm epoch in modified f(r) gravity consistent with solar system tests}, \href{https://doi.org/10.1016/j.physletb.2007.10.027}{\emph{Phys. Lett. B} {\bfseries 657} (2007) 238}.

\bibitem{Turner}
S.M.~{Carroll}, V.~{Duvvuri}, M.~{Trodden} and M.~{Turner}, \emph{Is cosmic speed-up due to new gravitational physics?}, \href{https://doi.org/10.1103/PhysRevD.70.043528}{\emph{Phys. Rev. D} {\bfseries 70} (2004) 043528}.

\bibitem{Nojiri-5}
S.~{Nojiri} and S.~{Odintsov}, \emph{Unified cosmic history in modified gravity: From f(r) theory to lorentz non-invariant models}, \href{https://doi.org/10.1016/j.physrep.2011.04.001}{\emph{Phys. Rep.} {\bfseries 505} (2011) 59}.

\bibitem{Nojiri-4}
S.~{Nojiri}, S.~{Odintsov} and V.~{Oikonomou}, \emph{Modified gravity theories on a nutshell: Inflation, bounce and late-time evolution}, \href{https://doi.org/10.1016/j.physrep.2017.06.001}{\emph{Phys. Rep.} {\bfseries 692} (2017) 1}.

\bibitem{Olmo}
G.~{Olmo}, D.~{Rubiera-Garcia} and A.~{Wojnar}, \emph{Stellar structure models in modified theories of gravity: Lessons and challenges}, \href{https://doi.org/https://doi.org/10.1016/j.physrep.2020.07.001}{\emph{Phys. Rep.} {\bfseries 876} (2020) 1}.

\bibitem{Paxton_2011}
B.~{Paxton}, L.~{Bildsten}, A.~{Dotter}, F.~{Herwig}, P.~{Lesaffre} and F.~{Timmes}, \emph{Modules for experiments in stellar astrophysics (mesa)}, \href{https://doi.org/10.1088/0067-0049/192/1/3}{\emph{Astrophys. J. Suppl.} {\bfseries 192} (2011) 3}.

\bibitem{Paxton_2013}
B.~{Paxton}, {Cantiello, M.}, {Arras, P.}, {Bildsten, L.}, {Brown, E.F.}, {Dotter, A.} et~al., \emph{Modules for experiments in stellar astrophysics (mesa): Planets, oscillations, rotation, and massive stars}, \href{https://doi.org/10.1088/0067-0049/208/1/4}{\emph{Astrophys. J. Suppl.} {\bfseries 208} (2013) 4}.

\bibitem{Paxton_2015}
B.~{Paxton}, {Marchant, P.}, {Schwab, J.}, {Bauer, E.B.}, {Bildsten, L.}, {Cantiello, M.} et~al., \emph{Modules for experiments in stellar astrophysics (mesa): Binaries, pulsations, and explosions}, \href{https://doi.org/10.1088/0067-0049/220/1/15}{\emph{Astrophys. J. Suppl.} {\bfseries 220} (2015) 15}.

\bibitem{Paxton_2018}
B.~{Paxton}, {Schwab, J.}, {Bauer, E.B.}, {Bildsten, L.}, {Blinnikov, S.}, {Duffell, P.} et~al., \emph{Modules for experiments in stellar astrophysics (mesa): Convective boundaries, element diffusion, and massive star explosions}, \href{https://doi.org/10.3847/1538-4365/aaa5a8}{\emph{Astrophys. J. Suppl.} {\bfseries 234} (2018) 34}.

\bibitem{Ohlmann_2017}
S.~{Ohlmann}, F.~{Röpke}, R.~{Pakmor} and V.~{Springel}, \emph{Constructing stable 3d hydrodynamical models of giant stars}, \href{https://doi.org/10.1051/0004-6361/201629692}{\emph{Astron. Astrophys.} {\bfseries 599} (2017) A5}.

\bibitem{Joyce_2019}
M.~{Joyce}, L.~{Lairmore}, D.J.~{Price}, S.~{Mohamed} and T.~{Reichardt}, \emph{Density conversion between 1d and 3d stellar models with $^{1D}$mesa2hydro$^{3D}$}, \href{https://doi.org/10.3847/1538-4357/ab3405}{\emph{Astrophys. J.} {\bfseries 882} (2019) 63}.

\bibitem{Gorski_2005}
K.M.~{G{\'o}rski}, E.~{Hivon}, A.J.~{Banday}, B.D.~{Wandelt}, F.K.~{Hansen}, M.~{Reinecke} et~al., \emph{Healpix: A framework for high-resolution discretization and fast analysis of data distributed on the sphere}, \href{https://doi.org/10.1086/427976}{\emph{Astrophys. J.} {\bfseries 622} (2005) 759}.

\bibitem{Pakmor_2012}
R.~{Pakmor}, P.~{Edelmann}, F.K.~{Röpke} and W.~{Hillebrandt}, \emph{Stellar gadget: a smoothed particle hydrodynamics code for stellar astrophysics and its application to type ia supernovae from white dwarf mergers}, \href{https://doi.org/10.1111/j.1365-2966.2012.21383.x}{\emph{MNRAS} {\bfseries 424} (2012) 2222}.

\bibitem{Rosswog_2004}
S.~{Rosswog}, R.~{Speith} and G.~{Wynn}, \emph{{Accretion dynamics in neutron star-black hole binaries}}, \href{https://doi.org/10.1111/j.1365-2966.2004.07865.x}{\emph{MNRAS} {\bfseries 351} (2004) 1121}.

\bibitem{Springel_2010}
V.~{Springel}, \emph{{E pur si muove: Galilean-invariant cosmological hydrodynamical simulations on a moving mesh}}, \href{https://doi.org/10.1111/j.1365-2966.2009.15715.x}{\emph{MNRAS} {\bfseries 401} (2010) 791}.

\bibitem{Pakmor_2016}
S.~{Ohlmann}, F.~{R{\"o}pke}, R.~{Pakmor}, V.~{Springel} and E.~{M{\"u}ller}, \emph{{Magnetic field amplification during the common envelope phase}}, \href{https://doi.org/10.1093/mnrasl/slw144}{\emph{MNRAS} {\bfseries 462} (2016) L121}.

\bibitem{Weinberger}
R.~{Weinberger}, V.~{Springel} and R.~{Pakmor}, \emph{The arepo public code release}, \href{https://doi.org/10.3847/1538-4365/ab908c}{\emph{Astrophys. J. Suppl.} {\bfseries 248} (2020) 32}.

\bibitem{Mukhopadhyay} B. {Mukhopadhyay}, \emph{Description of pseudo-newtonian potential for the relativistic accretion disk around kerr black holes}, \emph{Astrophys. J.} {\bfseries 581} (2002) 427.

\bibitem{Turk_2011}
M.~{Turk}, B.~{Smith}, J.~{Oishi}, S.~{Skory}, S.~{Skillman}, T.~{Abel} et~al., \emph{yt: A multi-code analysis toolkit for astrophysical simulation data}, \href{https://doi.org/10.1088/0067-0049/192/1/9}{\emph{Astrophys. J. Suppl.} {\bfseries 192} (2010) 9}.

\bibitem{Hunter}
J.D.~Hunter, \emph{Matplotlib: A 2d graphics environment}, \href{https://doi.org/10.1109/MCSE.2007.55}{\emph{Computing in Science \& Engineering} {\bfseries 9} (2007) 90}.

\end{thebibliography}\endgroup





\end{document}